\documentclass{article}

\usepackage[english]{babel}
\usepackage{amssymb}
\usepackage{authblk}
\usepackage[letterpaper,top=2cm,bottom=2cm,left=3cm,right=3cm,marginparwidth=1.75cm]{geometry}
\usepackage{comment}
\usepackage{amsmath}
\usepackage{graphicx}
\usepackage{xcolor}
\usepackage{multirow}
\usepackage[colorlinks=true, allcolors=blue]{hyperref}
\usepackage{booktabs}
\usepackage{makecell}
\usepackage{adjustbox}
\usepackage{tabularx}
\usepackage{array}
\usepackage{adjustbox}
\usepackage{hyperref}

\title{Reproducible capillary fluctuation analysis of solid–liquid interfaces for stiffness and anisotropy calculations}

\author[1,2]{Kai Liu}
\author[3]{Douglas E. Spearot}
\author[1,*]{Damien Tourret}

\affil[1]{IMDEA Materials, Getafe, 28906 Madrid, Spain}
\affil[2]{Scientific Instrumentation and Process Technology unit, Luxembourg Institute of Science and Technology (LIST),  Esch-sur-Alzette, Luxembourg}
\affil[3]{Department of Mechanical and Aerospace Engineering, University of Florida, Gainesville, FL, United States}

\affil[*]{Corresponding author: damien.tourret@imdea.org}

\date{}

\begin{document}
\maketitle

\begin{abstract}
The capillary fluctuation method (CFM) is widely used to compute solid--liquid interfacial properties from atomistic simulations, but its accuracy depends on choices in interface construction, wave-vector selection, sampling, and simulation geometry. Here, we develop a diagnostics-driven workflow for reproducible CFM calculations using pure Al as a representative system. Employing both ribbon models and thick two-dimensional references, we show that apparent linearity of the fluctuation spectrum alone does not ensure reliable stiffness or anisotropy estimates. Instead, a reliable CFM analysis requires a fitting window consistent with both temporal sampling and continuum capillary-wave assumptions, systematic sensitivity tests of the interface identification procedure, explicit propagation of replica variability, and independent verification of model-thickness convergence. We further propose a practical thickness-selection rule based on coexistence-temperature consistency, which enables finite-size effects to be controlled while retaining the substantial computational efficiency of ribbon geometries.
By making the main sources of uncertainty explicit and diagnosable, the proposed workflow improves the reliability of the CFM as a quantitative tool and provides a foundation for its broader application to complex solid--liquid interfaces. 
\end{abstract}

\section*{Introduction}
The free energy of a solid--liquid interface (SLI) and its anisotropy play a central role in the selection of solidification microstructures, crystal growth morphologies, and dendrite orientations~\cite{wang2020controlling,haxhimali2006orientation}. Although the anisotropy of the interfacial free energy is often small in metals and alloys, its effect on microstructural evolution is disproportionately large. Accurate quantification of this anisotropy is therefore essential for predictive models, including phase-field simulations~\cite{zhao2022role,tourret2022phase,karma2016atomistic}.

Among available atomistic approaches, the capillary fluctuation method (CFM) has become one of the most established and widely used techniques for evaluating SLI properties~\cite{hoyt2001method,asta2009solidification,ou2020imaging}. By analyzing the equilibrium thermal fluctuations of a rough interface, the CFM provides direct access to the interfacial stiffness, which contains both the interfacial free energy and its orientational derivatives. Since stiffness anisotropy is typically more pronounced than free-energy anisotropy, the CFM offers a particularly sensitive route for characterizing anisotropic interfacial behavior. 
CFM calculations can be performed using either ribbon geometries with a quasi-one-dimensional SLI or thicker simulation cells that retain the full two-dimensional SLI. The two representations are theoretically consistent and are expected to yield equivalent results when finite-size and sampling effects are properly controlled~\cite{du2007properties}. The CFM has consequently been applied to a wide range of SLIs, including metals with different crystal structures~\cite{sun2004crystal,asadi2015two,asadi2016anisotropy}, ionic compounds~\cite{benet2015interfacial}, ice--water interfaces~\cite{benet2014study,llombart2020surface}, and organic molecular systems~\cite{feng2006calculation,gerges2015predictive}.

The relevance of the CFM is further reinforced by recent developments in both simulation methodology and materials applications. Advances in atomistic simulation, particularly the development of machine-learned interatomic potentials (MLIP), have significantly improved the accuracy and chemical complexity accessible in interfacial calculations~\cite{mishin2021machine,jakse2023machine}. At the same time, emerging applications such as additive manufacturing~\cite{hua2024room}, metal recycling~\cite{kotadia2025aluminium}, and battery materials~\cite{yang2020dendrites} increasingly require accurate interfacial parameters for predicting microstructure evolution, including texture selection during rapid solidification~\cite{zhong2025quantification}. In parallel, the conceptual framework of the CFM has been extended beyond its traditional use for equilibrium rough SLIs to other fluctuating interfaces, including grain boundary stiffness~\cite{foiles2006computation}, nanograin stability~\cite{hussein2024model}, and colloidal nanoparticle assembly~\cite{ou2020imaging}. For faceted interfaces governed by step-flow growth, the CFM has also been shown to remain applicable within the step plane~\cite{liang2018plane}. A recent work has even extended the approach to moving interfaces~\cite{urashima2026capillary}. These developments highlight the CFM not only as a mature tool for equilibrium SLI characterization, but also as an increasingly versatile framework for interfacial thermodynamics across broader materials contexts.

Systematic studies of CFM accuracy and reliability remain surprisingly limited. Despite the maturity of the underlying theory, practical implementations still exhibit notable variability across the literature. Key methodological choices, such as the definition of the instantaneous interface~\cite{chacon2003intrinsic,partay2008new}, the selection of wave-vector ranges, the treatment of finite sampling, and the handling of finite-size effects, are often made empirically and are not reported in a sufficiently consistent way.
Furthermore, the choice of model thickness varies considerably across different implementations. Although some studies have conducted tests on thickness convergence~\cite{morris2002complete,ambler2017solid}, the consistency between quasi-1D interfaces (known as ribbon models) and 2D interfaces has not been fully verified~\cite{benet2015interfacial,rozas2011capillary}.
These gaps matter because CFM results are frequently used to interpret subtle physical effects, including the influence of composition~\cite{ueno2019composition,azizi2022interactive}, undercooling~\cite{yin2024far}, strain~\cite{yin2024far}, and thermal gradients~\cite{brown2017interfacial,swamy2026unfolding} on interfacial properties. When the physical effect of interest is small, poorly quantified CFM uncertainty can lead to misleading trends or apparent disagreements between studies.

Thus, the objective of this work is to develop and validate a diagnostics-driven workflow for reproducible CFM analysis of SLIs, providing a methodological basis for broader applications of the CFM. Using pure Al as a representative system, we construct a hierarchy of ribbon and thick 2D models to examine the effects of simulation geometry and interface representation. We first establish a reliable capillary-wave fitting regime by combining relaxation-time analysis with sensitivity tests of interface construction, coarse graining, and fitting-window selection. We then assess the convergence of interfacial stiffness with model thickness and compare the conventional 1D ribbon model with two different approaches for the 2D SLI. Replica-based uncertainty propagation is further used to quantify the robustness of the resulting anisotropy parameters and their convergence toward the thick-model reference. Finally, we discuss practical guidelines for model-thickness selection and the broader implications of the proposed workflow for reproducible and computationally efficient CFM calculations.

\section*{Results}\label{sec:results}

First, it is essential for the interpretation of our results to recall some key theoretical concepts related to the CFM. 
For cubic crystals, the orientation dependence of the SLI free energy is commonly parameterized using a cubic harmonic expansion. A commonly used parameterization is the conventional two-parameter cubic harmonic expansion up to sixth order~\cite{Fehlner1976APR},
\begin{equation}
\gamma(\hat{\mathbf{n}})=\gamma_{0}
\left[
1+\varepsilon_{1}\left(\sum_{i=1}^{3} n_i^{4}-\frac{3}{5}\right)
+\varepsilon_{2}\left(
3\sum_{i=1}^{3} n_i^{4}
+66 n_1^2 n_2^2 n_3^2
-\frac{17}{7}
\right)
\right],
\label{Eq:cubic_harmonic_expansion}
\end{equation}
where $\gamma_0$ is the orientation-averaged interfacial free energy, $\varepsilon_1$ and $\varepsilon_2$ are anisotropy parameters, and $\hat{\mathbf{n}}=\langle n_1,n_2,n_3\rangle$ is the interface normal. The overarching aim of the CFM is thus to calculate parameters $\gamma_0$, $\varepsilon_1$, and $\varepsilon_2$.

Following the seminal work by Hoyt \textit{et al.}~\cite{hoyt2001method}, instead of directly estimating the excess free energy, $\gamma$, as a function of the interface orientation, $\theta$, one may rather extract the interfacial stiffness
\begin{equation}
\tilde{\gamma}(\theta) = \gamma(\theta) + \frac{d^2 \gamma(\theta)}{d\theta^2},
\end{equation}
whose anisotropy is significantly stronger compared to that of $\gamma$.

The CFM is based on analyzing the thermal fluctuations of a SLI at equilibrium. 
Consider a general SLI described as a two-dimensional height field $h(x,y)$ over an interfacial area $A=L_xL_y$. Under the assumptions of (i) a continuum description of the interface, (ii) a sharp interface whose thickness is negligible compared to the wavelength of fluctuations, and (iii) small interfacial slopes ($|\nabla h| \ll 1$), the free energy can be expanded to quadratic order in terms of the in-plane interfacial-stiffness tensor $\boldsymbol{\Gamma}$~\cite{hoyt2001method,du2007properties} as
\begin{equation}
F \approx \frac{1}{2}\int_A (\nabla h)^{\mathrm T}\boldsymbol{\Gamma}(\nabla h)\,dA.
\end{equation}
For a Fourier mode with $\mathbf{k}=(k_x,k_y)$, equipartition gives
\begin{equation}
\frac{k_BT}{L_xL_y\left\langle |A(\mathbf{k})|^2\right\rangle}
=\mathbf{k}^{\mathrm T}\boldsymbol{\Gamma}\mathbf{k}
=\Gamma_{xx}k_x^2+2\Gamma_{xy}k_xk_y+\Gamma_{yy}k_y^2,
\label{Eq:stiffness_tensor_fit_equation}
\end{equation}
in which $A(\mathbf{k})$ is the Fourier amplitude of the height profile~\cite{becker2009atomistic}. 
A sufficiently thick model for which nonzero $k_y$ modes fall within the accepted low-$k$ window can therefore be analyzed by a full 2D tensor fit. 

For a ribbon model which exhibits a quasi-1D SLI described by the height field $h(x)$, Eq.~\eqref{Eq:stiffness_tensor_fit_equation} reduces to
\begin{equation}
\frac{k_BT}{L_xL_y\left\langle |A(k)|^2\right\rangle}=\Gamma_{xx}k_x^2.
\label{Eq:stiffness_fit_equation_revised}
\end{equation}
The ribbon model gives rise to larger amplitude fluctuations than systems with 2D interfaces~\cite{karma1993fluctuations}. For the same minimum wave number $k_{min}$, the 1D model requires less computational effort.

It is important to note that the validity of this approach relies on the assumptions outlined above. In atomistic simulations, these assumptions are not strictly satisfied, which can lead to systematic deviations in the extracted interfacial stiffness.

In addition to their equilibrium amplitudes, capillary modes also exhibit characteristic relaxation dynamics. 
Following the fluctuation-spectrum framework of Hoyt \textit{et al.}~\cite{hoyt2002atomistic}, the interface height evolves through curvature-driven relaxation in the presence of thermal noise. 
For a Fourier mode \(A(k,t)\), the deterministic part of this dynamics can be written as
\begin{equation}
\frac{d A(k,t)}{dt}
=
-\mu \tilde{\gamma} k^2 A(k,t),
\end{equation}
where \(\mu\) is the kinetic coefficient of the SLI and \(\tilde{\gamma}\) is the interfacial stiffness. 
The factor \(k^2\) arises because the capillary driving force is proportional to the local curvature of the interface.

This equation implies an exponential relaxation of each Fourier mode,
\begin{equation}
\left\langle A(k,t) A^*(k,0) \right\rangle
\propto
\exp\left[-\frac{t}{\tau(k)}\right],
\end{equation}
with the relaxation time given by
\begin{equation}
\tau(k)
=
\frac{1}{\mu \tilde{\gamma} k^2}.
\label{Eq:relaxation time}
\end{equation}

\subsection*{Diagnostics-driven workflow for reproducible CFM analysis}

\begin{figure}[t!]
    \centering
    \includegraphics[width=1\linewidth]{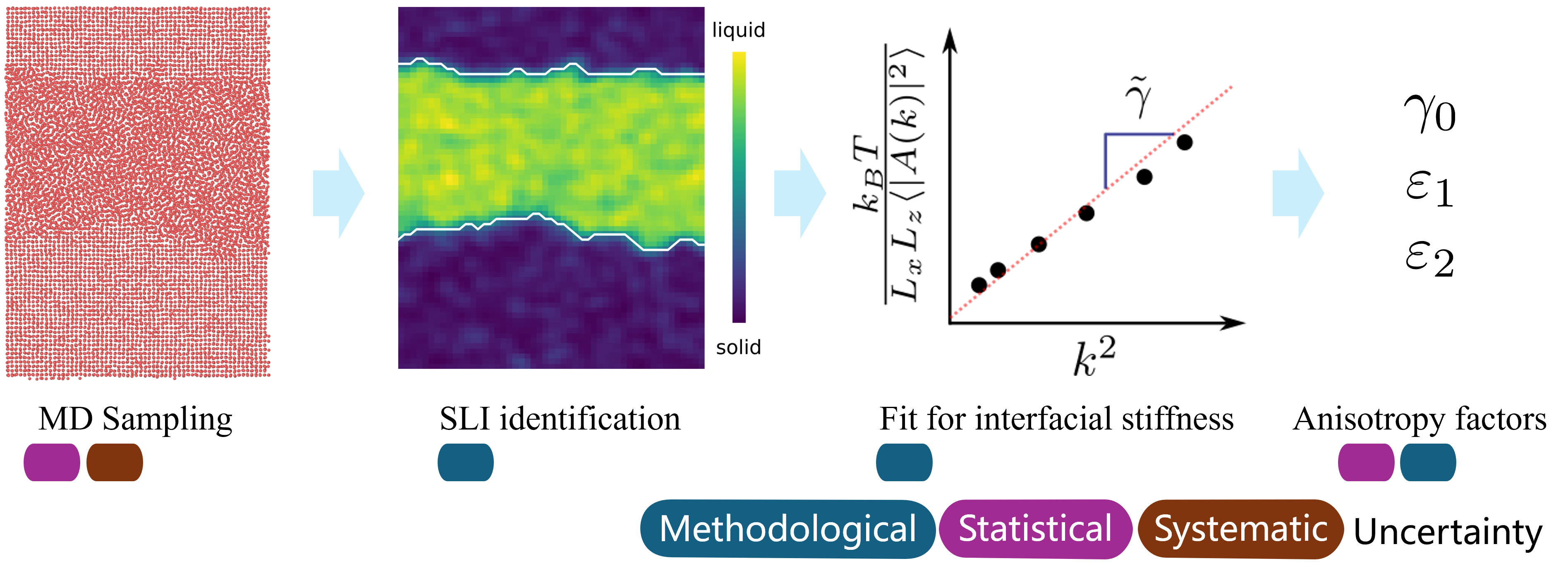}
    \caption{Workflow and uncertainty sources in CFM calculations of solid--liquid interfacial properties. The calculation proceeds through four stages: MD sampling, SLI identification, stiffness fitting, and anisotropy fitting. The color bands indicate the dominant uncertainty sources associated with each stage. Methodological uncertainty mainly arises from interface construction and fitting-window selection, statistical uncertainty from finite-time sampling of capillary modes, and systematic uncertainty from finite simulation geometry, coexistence condition, residual lateral stress, and model thickness.
    }
    \label{fig:CFM_workflow}
\end{figure}

In practice, a CFM calculation involves four consecutive steps: 
(i) molecular dynamics (MD) sampling of an equilibrated two-phase system, 
(ii) construction of an instantaneous SLI from atomistic configurations, 
(iii) fitting of the fluctuation spectrum to obtain orientation-dependent stiffnesses, 
and (iv) conversion of these stiffnesses into anisotropy parameters \(\gamma_0\), \(\varepsilon_1\), and \(\varepsilon_2\).
Each step involves methodological or simulation-design choices that can affect the final stiffness and anisotropy estimates. 
As summarized schematically in Fig.~\ref{fig:CFM_workflow}, these uncertainty sources are not confined to a single stage of the calculation, but can propagate through the entire workflow.

For clarity, the uncertainty sources discussed in this work are classified into three categories: methodological uncertainty arising from analysis choices, statistical uncertainty associated with finite sampling, and systematic uncertainty associated with finite simulation geometry and the resulting coexistence behavior.
Distinguishing these sources of uncertainty helps clarify how they can be diagnosed, controlled, and reported in reproducible CFM calculations.

Reliable CFM calculations require consistency checks across the entire workflow. 
Table~\ref{tab:cfm_diagnostic_summary} summarizes the main uncertainty sources considered in the present work, together with the corresponding diagnostics and validation strategies. 
The purpose of this workflow is not to prescribe universal numerical parameters, but to provide a practical map of the quantities that should be examined and reported in reproducible CFM studies. 

\begin{figure}
    \centering
    \includegraphics[width=0.75\linewidth]{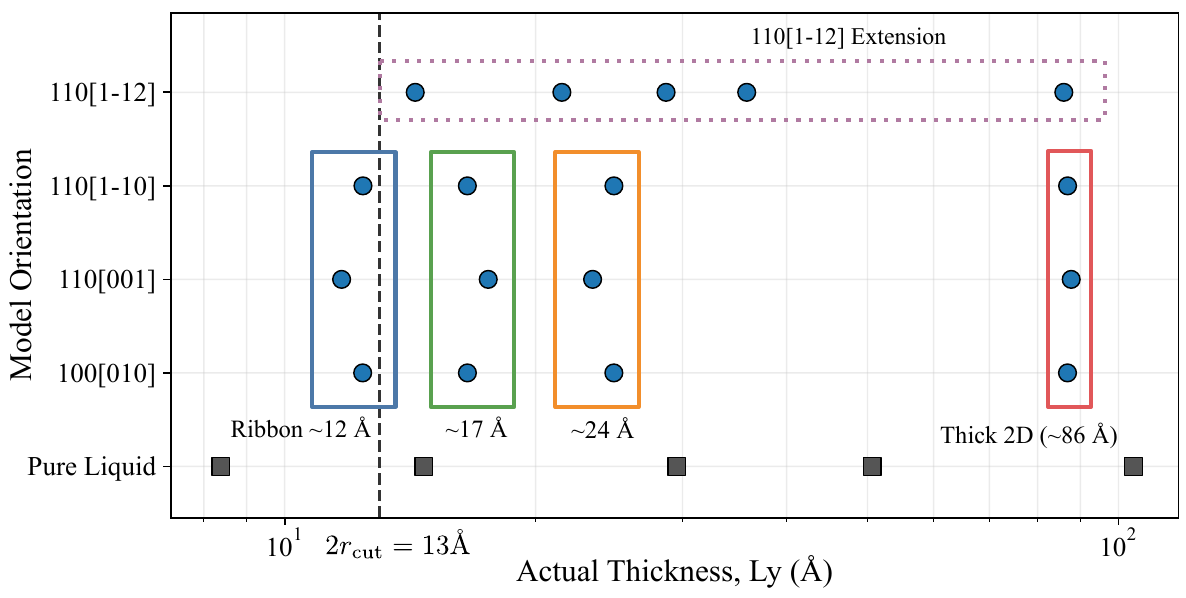}
    \caption{
Overview of the simulation models used in this work, including ribbon models with varied thicknesses and thick 2D reference models. Models with comparable thicknesses are enclosed by solid outlines, and the corresponding interfacial stiffnesses are used to determine the interfacial anisotropy parameters. 110[1\(\bar{1}\)2] models are used for convergence diagnostic. Pure-liquid slabs are included to examine the thickness dependence of liquid-phase properties. Each SLI data point represents three independent replicas. The dashed line marks \(2r_{\mathrm{cut}}=13\)~\AA{}.
}
    \label{fig:model_thickness_design}
\end{figure}

These diagnostics are demonstrated using pure Al as a representative test system.
The simulation cells were designed as a hierarchy of models to assess finite-thickness effects and to establish a reliable reference for the CFM analysis, as summarized in Fig.~\ref{fig:model_thickness_design}. 
For the three main SLI orientations, ribbon models with thicknesses of approximately 12, 17, and 24~\AA{} were constructed to examine convergence with respect to the transverse dimension \(L_y\). 
Additionally, a series of 110[1\(\bar{1}\)2] models were included for thickness convergence check. 
Much thicker models with \(L_y\approx86\)~\AA{} were used as two-dimensional reference systems, for which the full interfacial-stiffness tensor can be determined. 
In parallel, pure-liquid models spanning a wider range of \(L_y\) were simulated to characterize finite-size effects in the liquid phase itself. 
The dashed line in Fig.~\ref{fig:model_thickness_design} indicates \(2r_{\mathrm{cut}}=13\)~\AA{}, emphasizing that exceeding twice the interaction cutoff provides only a lower geometric bound rather than a convergence criterion. Further simulation details are provided in the Methods section.

\begin{table*}[h!]
\centering
\caption{
Summary of key diagnostics used in the proposed CFM workflow.
}
\label{tab:cfm_diagnostic_summary}
\renewcommand{\arraystretch}{1.25}
\begin{tabularx}{\textwidth}{
    >{\raggedright\arraybackslash}p{3.2cm}
    >{\raggedright\arraybackslash}X
    >{\raggedright\arraybackslash}X
}
\toprule
\textbf{Workflow}
& \textbf{Main diagnostic}
& \textbf{Purpose} \\
\midrule

MD sampling
& Solid-fraction stability, residual-stress reporting, and independent replicas
& Verify stable coexistence and trajectory variability \\

\midrule

\(k\)-window selection
& Relaxation-time analysis and \(k\)-window sensitivity
& Select temporally and spatially valid fluctuation modes \\

\midrule

Interface construction
& Descriptor, coarse-graining, and interface-extraction sensitivity
& Test robustness of the constructed interface \\

\midrule

Interface representation
& Comparison of 1D full-\(Y\), 2D \(k_y=0\), and full 2D tensor analyses
& Assess sensitivity to the treatment of transverse interfacial fluctuations \\

\midrule

Sampling convergence
& Sampling-duration convergence
& Check finite-time convergence of extracted stiffness\\

\midrule

Thickness convergence
& Equilibrium-temperature and stiffness convergence
& Check transverse finite-size effects \\

\midrule

Liquid-phase finite-size check
& Density, directional RDF, and residual stress as functions of \(L_y\)
& Check transverse finite-size effects in the liquid phase \\
\midrule

Thick-2D reference
& Ribbon \(k_y=0\) vs. thick-2D \(k_y=0\) and full tensor
& Benchmark finite-thickness and analysis effects \\

\midrule

Anisotropy fitting
& Replica-based uncertainty propagation
& Propagate uncertainty and test consistency \\

\bottomrule
\end{tabularx}
\end{table*}

\subsection*{Relaxation dynamics constrain the accessible capillary-wave modes}
\label{sec:k_relaxation_diagnostics}

\begin{figure}[b!]
    \centering
    \includegraphics[width=\textwidth]{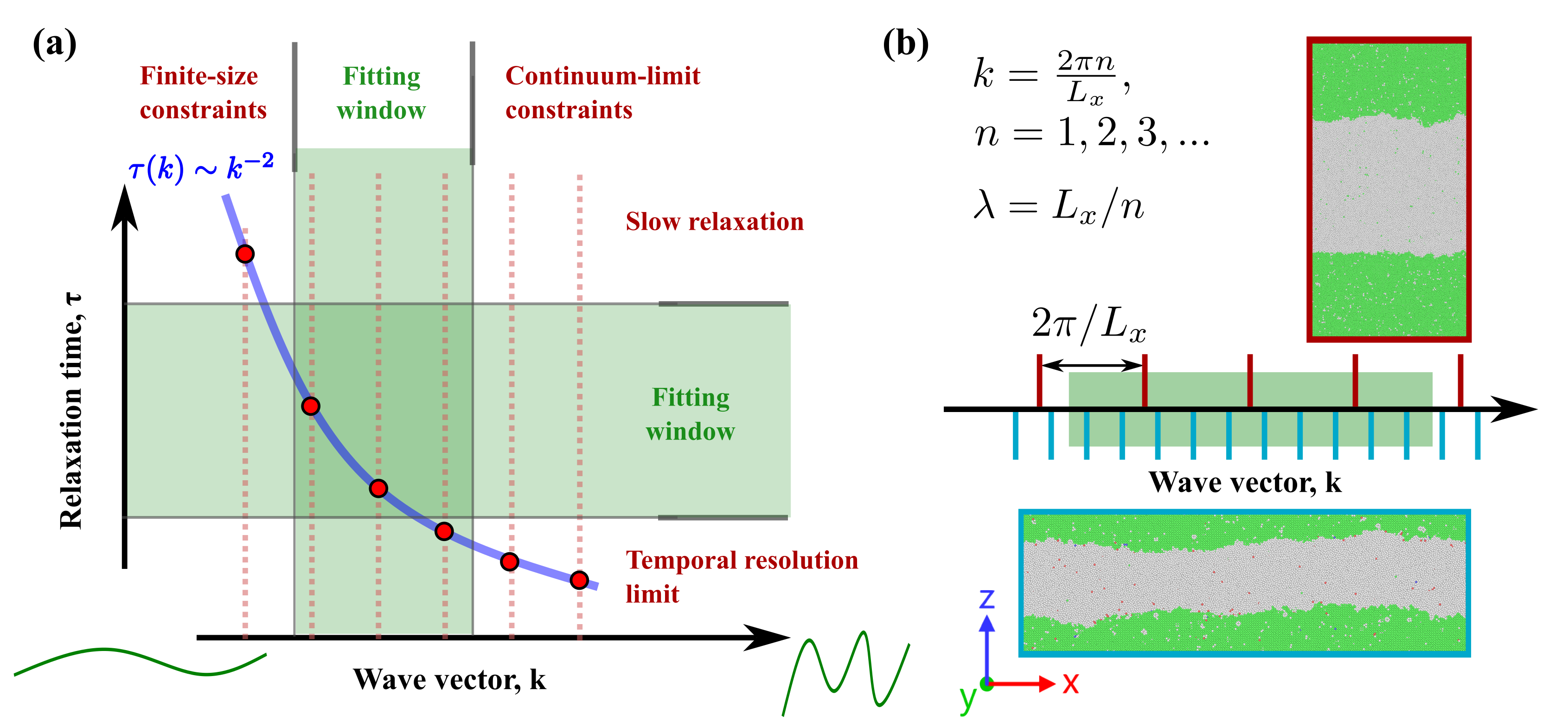}
    \caption{
    Spatial and temporal constraints on the selection of the fitting wave-vector window in CFM.
    (a) The usable \(k\)-window is bounded at low \(k\) by finite-size and slow-relaxation effects, and at high \(k\) by the breakdown of the continuum capillary-wave description and by temporal-resolution limits.
    The relaxation time follows approximately \(\tau(k)\sim k^{-2}\), linking the spatial wavelength of a mode to the sampling time required to resolve it.
    (b) The accessible wave vectors are discretized by the interfacial length \(L_x\), \(k_n=2\pi n/L_x\).
    For the same physically meaningful \(k\)-window, a larger \(L_x\) provides a denser set of accessible modes and can therefore improve the reliability of stiffness fitting.
    }
    \label{fig:k_window_selection}
\end{figure}

In many CFM studies, the fitting range is selected mainly from the apparent linearity of the fluctuation spectrum, often judged from log--log plots~\cite{hoyt2001method,asta2002calculation}. 
However, such a criterion is inherently subjective and does not ensure that the retained modes remain within the continuum capillary-wave regime or are sufficiently sampled in time. 
A reliable fitting window must satisfy both spatial and temporal constraints, as illustrated in Fig.~\ref{fig:k_window_selection}(a).

For a MD model with interfacial length \(L_x\), the accessible wave vectors are discrete,
\begin{equation}
k_n = \frac{2\pi n}{L_x}, \quad n=1,2,3,\ldots.
\label{Eq:Lx_k_selection}
\end{equation}
At low \(k\), the wavelength is long and the relaxation time increases proportionally to $k^{-2}$, as
\begin{equation}
\tau(k)
=
\frac{1}{\mu \tilde{\gamma} k^2}.
\label{Eq:relaxation time}
\end{equation}
Long-wavelength modes may provide too few statistically independent samples within a finite trajectory. At high \(k\), the wavelength approaches the SLI thickness and the length scales introduced by interface construction, so the extracted spectrum can be affected by the breakdown of the sharp-interface continuum approximation, atomistic noise, and smoothing artifacts. High-\(k\) modes also require a sufficiently short output interval to resolve their relaxation dynamics.

These considerations imply that the fitting window is not an independent post-processing parameter. Instead, \(L_x\), the simulation duration, the output interval, the interface-construction length scale, and the selected \(k\)-window must be designed together. 

\begin{figure}[t!]
    \centering
    \includegraphics[width=0.95\textwidth]{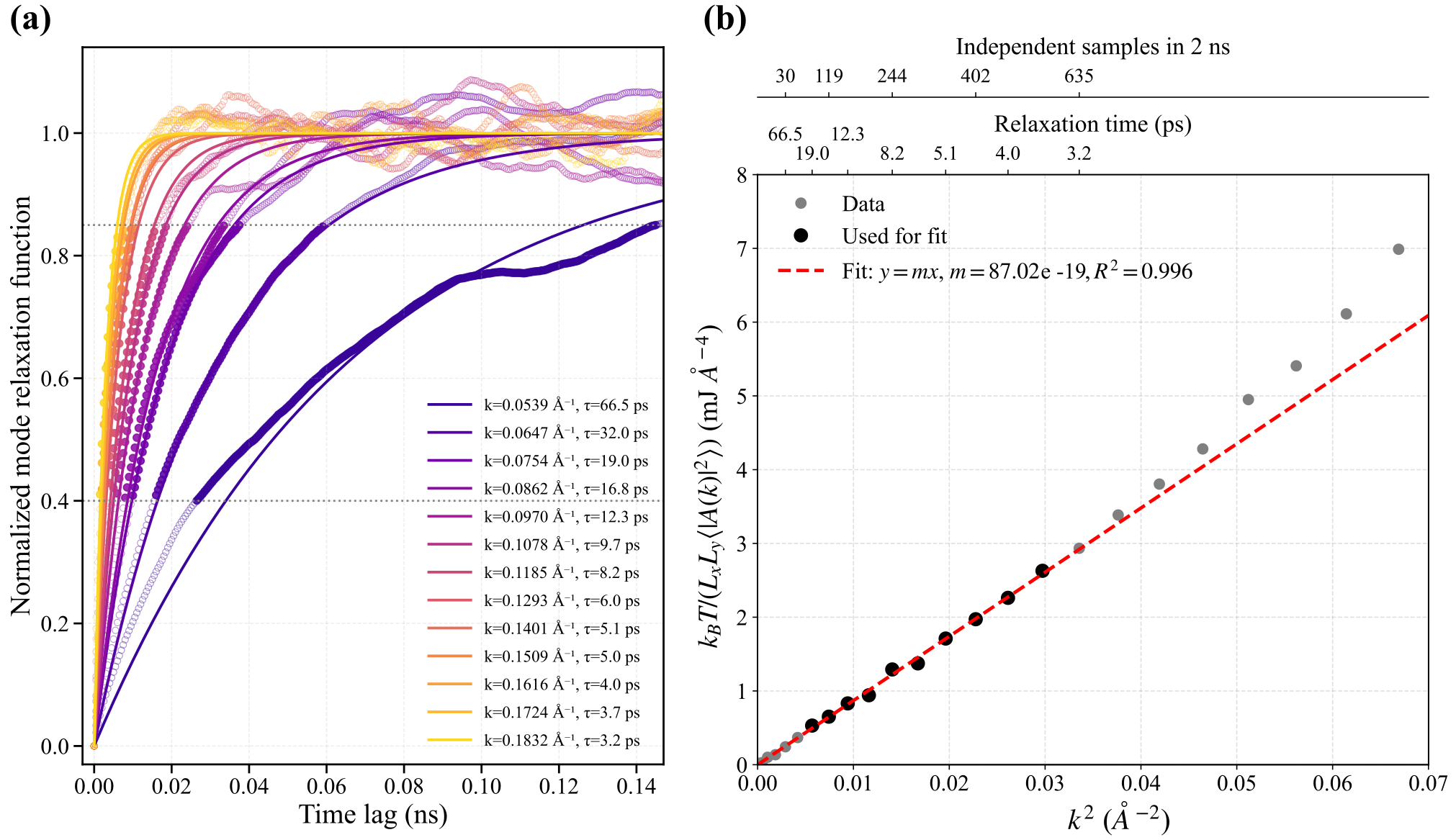}
    \caption{
    Relaxation diagnostics and fluctuation spectrum for the \(100[010]\) Al solid--liquid interface (24~\AA{} model with LOP descriptor and 2D \(k_y=0\) method). 
    (a) Mode-dependent relaxation times obtained by fitting the autocorrelation function of each Fourier mode to an exponential decay. The lowest wave vector, \(k=0.0539~\mathrm{\AA^{-1}}\), has a large fitting error and is excluded from the stiffness fit, whereas the intermediate-\(k\) modes show well-defined relaxation. 
    (b) Capillary fluctuation spectrum plotted as \(k_B T/[L_xL_y\langle |A(k)|^2\rangle]\) against \(k^2\), so that the slope of a zero-intercept linear fit gives the stiffness. The upper axis reports the corresponding relaxation time and the estimated number of independent samples in a \(5~\mathrm{ns}\) trajectory. The selected fitting range, \(0.005<k^2<0.03~\mathrm{\AA^{-2}}\), gives \(\tilde{\gamma}=87.0~\mathrm{mJ\,m^{-2}}\).}
    \label{fig:k_relaxation_spectrum}
\end{figure}

A CFM analysis is first performed for the 100[010] interface to evaluate the temporal sampling quality of the accessible capillary modes.
For each Fourier mode, the normalized autocorrelation function is computed and fitted to an exponential decay to estimate the relaxation time \(\tau(k)\), as shown in Fig.~\ref{fig:k_relaxation_spectrum}(a). 
The fitting is restricted to an intermediate correlation range, $0.4 < C(t)/C(0) < 0.85$, where the signal remains sufficiently above the statistical noise level while avoiding short-time deviations from ideal exponential relaxation caused by high-frequency fluctuations and interface-identification noise.

The relaxation analysis provides a practical diagnostic for identifying modes that are sufficiently resolved under the present sampling protocol. 
With a sampling interval of \(0.5~\mathrm{ps}\), modes with relaxation times shorter than a few picoseconds are difficult to resolve reliably [yellow dots in panel (a)]. 
As a practical criterion, at least 5--10 sampled points are required over the relaxation time scale for a meaningful exponential fit. 
Therefore, modes with \(\tau(k)\ge 3~\mathrm{ps}\) are considered temporally resolvable under the present sampling protocol. 
The smallest accessible mode, \(k=0.0539~\mathrm{\AA^{-1}}\), has a large uncertainty in its fitted relaxation time, indicating insufficient sampling of the corresponding long-wavelength fluctuation. 
This mode is therefore excluded from further consideration. 
In contrast, the intermediate-\(k\) modes show more robust exponential relaxation and provide a temporally reliable basis for stiffness extraction.

Fig.~\ref{fig:k_relaxation_spectrum}(b) shows the corresponding stiffness fitting. 
The upper axes report the relaxation time of each mode and the estimated number of independent samples in a \(2~\mathrm{ns}\) trajectory, calculated from \(t_{\mathrm{run}}/\tau(k)\). 
Although the fluctuation spectrum appears highly linear over a broad \(k\)-range, some modes that lie on the same apparent linear trend may still be poorly sampled according to the relaxation-time analysis. 
This demonstrates that regression quality or apparent linearity alone is not sufficient for selecting a reliable fitting window.

The relaxation-time analysis alone, however, does not determine whether a mode remains within the continuum capillary-wave regime. 
Therefore, the temporally resolvable range identified here is treated as an upper-level candidate range. 
The final fitting window is determined by additionally requiring stability of the extracted stiffness and anisotropy parameters with respect to interface construction and \(k\)-range selection.

\subsection*{Interface construction and fitting-window sensitivity reveal high-\(k\) bias}
\label{sec:methodological_sensitivity}

\begin{figure}[t!]
    \centering
    \includegraphics[width=0.95\linewidth]{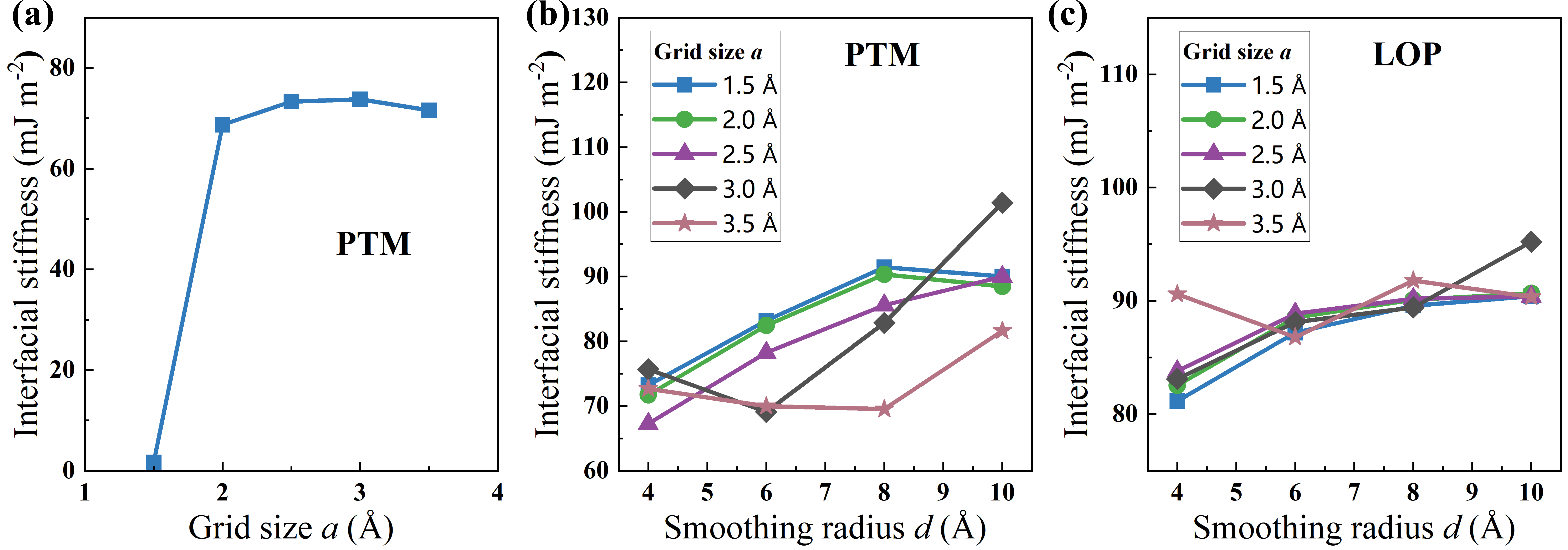}
\caption{
Sensitivity of the fitted interfacial stiffness to the coarse-graining parameters used to construct the continuous order-parameter field.
Results are compared for LOP- and PTM-based descriptors combined with averaging/smoothing over different radii and grid spacings: 
(a) PTM with bin averaging,
(b) PTM with kernel smoothing, and 
(c) LOP with kernel smoothing.
For the LOP descriptor, the fitted stiffness becomes nearly converged for a smoothing radius of \(d=6~\mathrm{\AA}\) and grid spacing below approximately \(2.5~\mathrm{\AA}\).
Smaller smoothing radii retain stronger sensitivity to the grid spacing, indicating incomplete suppression of local atomic-scale noise.
PTM-based coarse graining exhibits weaker convergence, while bin averaging becomes unreliable when the bin size is too small because too few atoms contribute to each spatial bin.
These results motivate the use of LOP with kernel smoothing, \(d=6~\mathrm{\AA}\), and a grid spacing of \(2.5~\mathrm{\AA}\) in the production analysis.
}
\label{fig:S_smoothing_parameter_selection}
\end{figure}

The second requirement for a reliable fitting window is that the retained modes remain within the regime where the constructed interface behaves as a continuum capillary wave. This condition is tested by examining the sensitivity of the fitted stiffness and anisotropy parameters to interface construction and fitting wave-vector range. In particular, we compare two distinct atomistic structural descriptors used to identify the two phases: the local order parameter (LOP)~\cite{hoyt2001method} and polyhedral template matching (PTM)~\cite{larsen2016robust}. The algorithmic details underlying these interface definitions are provided in the Methods section. 
The sensitivity of the fitted stiffness to the interface definition is evaluated for the \(100[010]\) (24~\AA{}) interface using different coarse-graining parameters and progressively enlarged \(k\)-windows. Here, \(a\) denotes the grid size and \(d\) denotes the smoothing radius, as defined in the Methods section.
The results are shown in Fig.~\ref{fig:S_smoothing_parameter_selection}. 
This preliminary test compares three interface-construction strategies applied to the same atomic configurations: PTM with bin averaging, PTM with kernel smoothing, and LOP with kernel smoothing.
For this screening step, the stiffness is obtained from the fitting range that gives the highest \(R^2\), without imposing the final \(k\)-window criterion.
Bin averaging becomes unreliable when \(a\) is too small, whereas kernel smoothing permits smaller grid spacings.
For kernel smoothing, the fitted stiffness depends more strongly on \(d\) than on \(a\).
A relatively stable parameter region is observed for the LOP-based construction with \(a \leq 3.0~\mathrm{\AA}\) and \(d \leq 6~\mathrm{\AA}\), but not for the PTM-based construction.
These screened parameters are used as candidates in the subsequent sensitivity analysis of the fitting \(k\)-window.

\begin{figure}[t!]
    \centering
    \includegraphics[width=0.8\textwidth]{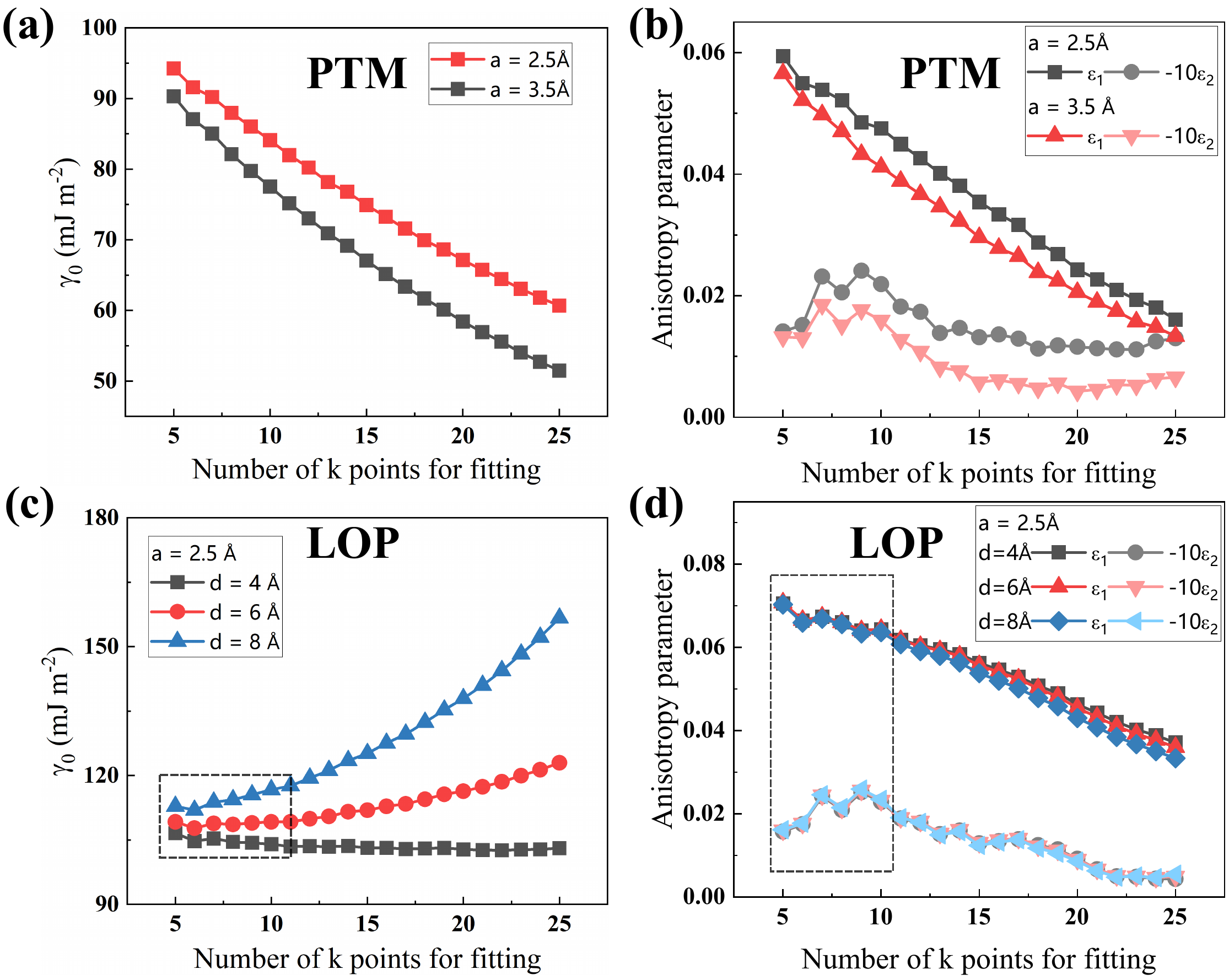}
    \caption{
Sensitivity of the fitted interfacial parameters to interface construction and
fitting wave-vector range.
Panels (a,c) show the orientation-averaged interfacial free energy \(\gamma_0\),
and panels (b,d) show \(\varepsilon_1\) and \(-10\varepsilon_2\), as functions
of the number of \(k\)-points used in the underlying stiffness fits.
Panels (a,b) use PTM with bin averaging, whereas panels (c,d) use LOP with
kernel smoothing.
The dashed box indicates the conservative low-\(k\) region in which the
LOP-based results become insensitive to the smoothing radius.
    }
    \label{fig:coarse_graining_sensitivity}
\end{figure}

Fig.~\ref{fig:coarse_graining_sensitivity} shows \(\gamma_0\) for the 24~\AA{} models (a,c) and the corresponding anisotropy parameters (b,d) obtained using different interface-construction procedures.
For each number of \(k\)-points, the three orientation-resolved stiffnesses are first fitted over the corresponding wave-vector range and are subsequently combined to determine \(\gamma_0\), \(\varepsilon_1\), and \(\varepsilon_2\). Panels (a,b) use PTM with bin averaging and the conventional 1D full-\(Y\) representation, following the commonly used ribbon-analysis procedure, whereas panels (c,d) use LOP with kernel smoothing and the 2D \(k_y=0\) representation (see Methods section).
The underlying stiffness fits retain high regression quality over the tested fitting ranges, with \(R^2>0.97\) for fits using 5 to 29 \(k\)-points.
Nevertheless, the fitted \(\gamma_0\) and anisotropy parameters show pronounced systematic variations, demonstrating that high regression quality alone does not guarantee robust CFM results.

For the PTM-based bin-averaging method, the fitted \(\gamma_0\) decreases continuously as more \(k\)-points are included in the stiffness fits, dropping from approximately \(90~\mathrm{mJ~m^{-2}}\) to below
\(60~\mathrm{mJ~m^{-2}}\). Changing the bin size from \(a=2.5~\mathrm{\AA}\) to \(a=3.5~\mathrm{\AA}\) does not remove this drift. The corresponding anisotropy parameters also vary systematically with the
number of fitted \(k\)-points, and the two bin sizes do not give consistent values even in the low-\(k\) region.

In contrast, the LOP-based kernel-smoothed interface shows a more stable low-\(k\) behavior.
Although \(\gamma_0\) still depends on the smoothing radius \(d\), especially when high-\(k\) modes are included, its low-\(k\) value becomes much less sensitive to the fitting range as \(d\) is reduced from
\(8~\mathrm{\AA}\) to \(4~\mathrm{\AA}\). More importantly, the anisotropy parameters obtained with different smoothing radii converge toward similar values in the low-\(k\) region.
For the present models, this stable region corresponds to approximately 7--10 \(k\)-points.
These observations support the use of the LOP-based kernel-smoothed interface for the subsequent CFM analysis.

The final fitting window is therefore chosen as the overlap between the temporally well-sampled range identified from relaxation diagnostics and the methodologically stable range identified from the interface-construction sensitivity test. This gives \(0.005 < k^2 < 0.03~\mathrm{\AA^{-2}}\),
corresponding to \(k=0.0755\)--\(0.1725~\mathrm{\AA^{-1}}\). For the \(100[010]\) interface of Al with the AEAM potential, a zero-intercept fit over this window gives \(\tilde{\gamma}=87.0~\mathrm{mJ\,m^{-2}}\), as shown in Fig.~\ref{fig:k_relaxation_spectrum}(b).

These results show that a coarse-graining method that appears acceptable for one fitting range can lead to biased \(\gamma_0\) and anisotropy parameters when shorter-wavelength modes are included. In the following analyses, the interface is therefore constructed using the LOP descriptor with kernel smoothing, a smoothing radius of \(d=6~\mathrm{\AA}\), and a grid spacing of \(2.5~\mathrm{\AA}\), and stiffnesses are fitted using the window defined above.

\subsection*{Model thickness and interface representation jointly control stiffness convergence}
\label{sec:thickness_convergence}

After establishing the interface-construction procedure and fitting \(k\)-window, we next examine the coupled effects of model thickness and interface representation. Previous CFM studies have shown that quasi-1D ribbon models can yield stiffnesses consistent with fully two-dimensional interfaces~\cite{du2007properties}, but this equivalence requires explicit verification when quantitative accuracy is sought. We therefore compare the 1D full-\(Y\), 2D \(k_y=0\), and full 2D tensor representations defined in the Methods section. The 2D \(k_y=0\) approach is of particular interest because it retains the computational efficiency of ribbon geometries while preserving transverse information during interface construction.

\begin{figure}
    \centering
    \includegraphics[width=1\linewidth]{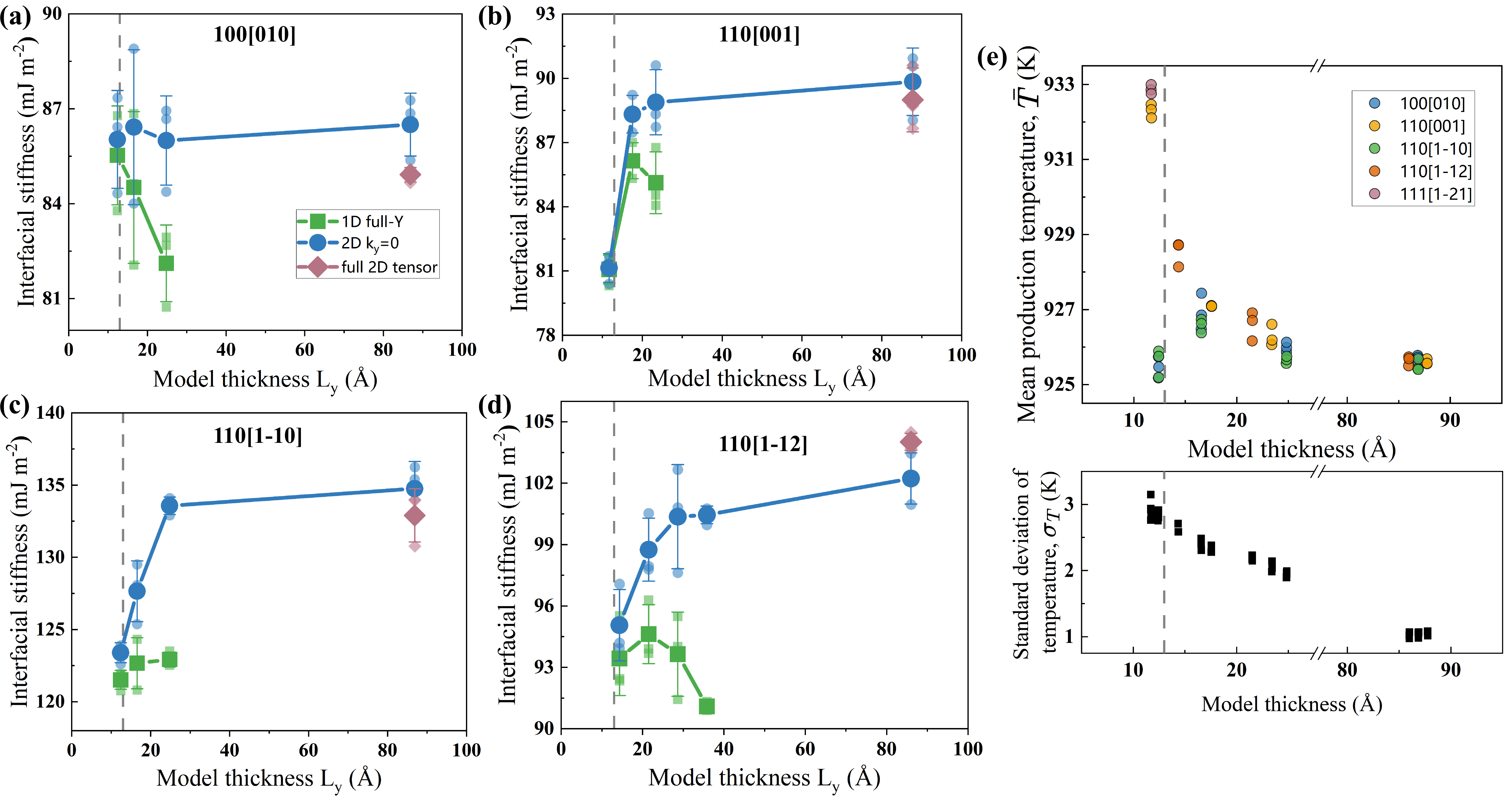}
    \caption{
Thickness dependence of the calculated interfacial stiffness and solid--liquid coexistence temperature. 
Panels (a--d) show the interfacial stiffnesses for the 100[010], 110[001], 110[1$\bar{1}$0], and 110[1$\bar{1}$2] orientations, respectively, using the model thicknesses summarized in Fig.~\ref{fig:model_thickness_design}. 
Green squares, blue circles, and purple diamonds denote the 1D full-\(Y\), 2D \(k_y=0\), and full 2D tensor representations, respectively. 
For each model, three independent replicas were performed; light-colored symbols show individual replicas, while dark-colored symbols indicate their mean values. Lines connecting the mean stiffness values are guides to the eye.
Panel (e) shows the mean equilibrium temperature sampled by each SLI model as a function of model thickness, with each symbol representing an independent replica. The lower panel reports the corresponding temporal standard deviation of the temperature, \(\sigma_T\), characterizing temperature fluctuations during sampling. A thin 111[1\(\bar{2}\)1] model is included for temperature diagnostic.
The gray dashed line marks \(2r_{\mathrm{cut}}=13\)~\AA{}.
}
    \label{fig:thickness_convergence}
\end{figure}

Thickness series were examined for the three orientations used in the primary anisotropy fit, namely 100[010], 110[001], and 110[1\(\bar{1}\)0], together with the additional 110[1\(\bar{1}\)2] orientation. All stiffnesses discussed here were obtained from 2~ns production trajectories, with three independent replicas for each SLI model. The results are shown in Fig.~\ref{fig:thickness_convergence}.

The 1D full-\(Y\) results show a pronounced sensitivity to model thickness, and the direction of the thickness dependence differs among orientations. No systematic convergence toward the thick-model results is observed. Such thickness sensitivity of interfaces obtained by averaging over the transverse direction has also been emphasized previously~\cite{ambler2017solid}. In contrast, the 2D \(k_y=0\) stiffnesses exhibit clear convergence with increasing \(L_y\). For the 100[010], 110[001], and 110[1\(\bar{1}\)0] orientations, the stiffnesses obtained from the \(\sim24\)~\AA{} ribbon models differ from the corresponding \(\sim86\)~\AA{} models by only 0.6\%, 1.1\%, and 0.9\%, respectively. The 110[1\(\bar{1}\)2] orientation converges more slowly, with a remaining difference of approximately 1.8\%.

For the thick 2D models, the stiffnesses obtained from the \(k_y=0\) modes also agree closely with those obtained from the full 2D tensor analysis, with differences below 2\% for all orientations. This agreement indicates that, once the transverse dimension is sufficiently large, the different interface representations converge toward a consistent stiffness.

For the three orientations used in the primary anisotropy fit, a thickness of approximately 24~\AA{} therefore provides stiffnesses already within about 1\% of the thick-model values when the 2D \(k_y=0\) representation is used. Importantly, the observed thickness dependence extends well beyond the direct interaction range of the interatomic potential (gray dashed lines). Significant changes in the extracted stiffness remain evident for models thicker than this limit, demonstrating that \(L_y>2r_{\mathrm{cut}}\) should be regarded only as a lower geometric bound rather than as a criterion for thickness convergence.

An additional indication of finite-thickness effects is provided by the solid--liquid coexistence temperature. All two-phase models were simulated using the same MD protocol, under which the temperature evolves toward the equilibrium coexistence condition in the \(NP_zAH\) ensemble (see Methods). The resulting equilibrium temperatures therefore provide an independent measure of thickness-dependent changes in the two-phase system. Figure~\ref{fig:thickness_convergence}(e) shows the mean production temperature for each independent simulation. For the thinnest models, with \(L_y\sim12\)~\AA{}, the equilibrium temperatures span a broad range of approximately 925--933~K among different orientations. Similar orientation-dependent coexistence temperatures have been reported in previous CFM studies~\cite{asadi2015two,yin2024far}. With increasing model thickness, however, the temperatures progressively converge. For the \(\sim86\)~\AA{} models, the four orientations, comprising 12 independent simulations in total, give an average coexistence temperature of \(925.6\pm0.1\)~K.

The lower part of Fig.~\ref{fig:thickness_convergence}(e) shows the temporal standard deviation of the instantaneous temperature during production sampling. As expected for a finite system, the magnitude of these fluctuations decreases approximately as
\begin{equation}
\sigma_T^{\mathrm{inst}} \propto L_y^{-1/2},
\end{equation}
consistent with the increase in the number of atoms as \(L_y\) increases. This statistical size effect accounts for the reduced amplitude of instantaneous temperature fluctuations in thicker models, but not for the systematic shift in their mean coexistence temperatures. The convergence of the mean temperature therefore reflects an additional finite-size effect beyond the expected reduction in sampling fluctuations.

We also examined pure-liquid models over a range of transverse thicknesses by monitoring quantities including density, pressure anisotropy, and directional RDF differences. Although these properties exhibit some thickness dependence, no single quantity provides a clear criterion for selecting the ribbon thickness, and several trends are entangled with conventional finite-size effects in MD sampling. The corresponding results are provided in the Supplementary Information S1, while the coexistence temperature therefore serves as a more direct diagnostic for the two-phase CFM system.

\subsection*{Replica-based uncertainty propagation affects anisotropy parameters}
\label{sec:statistical_uncertainty}

Even after the fitting wave-vector range and interface construction procedure are fixed, CFM results can retain residual variability due to finite sampling time and trajectory-to-trajectory differences. To quantify this effect, three independent simulations are performed for each interface orientation. The replicas use different random velocity seeds and slightly different initial liquid-slab positions while keeping the liquid volume fraction unchanged. The resulting coexistence temperatures, residual lateral stresses, and fitted stiffnesses are summarized in Table~\ref{tab:cfm_results_24A_ky0}.

\begin{table}[t]
\centering
\small
\caption{
Summary of thermodynamic conditions and 2D \(k_y=0\) CFM-derived stiffness for each independent replica in the nominal 24~\AA{} ribbon series. Temperature and pressure are reported as the mean $\pm$ temporal sample standard deviation over the production window (2\(~\mathrm{ns}\)). The mean stiffness is averaged over the three replicas of each orientation. The thickness of $110[1\bar{1}2]$ is $21.50$~\AA{}).
}
\label{tab:cfm_results_24A_ky0}
\begin{adjustbox}{max width=\textwidth}
\begin{tabular}{llccccc}
    \toprule
    Orientation & Replica &
    \makecell{Temperature (K)} &
    \makecell{\(P_{xx}\) (bar)} &
    \makecell{\(P_{yy}\) (bar)} &
    Stiffness (\(\mathrm{mJ\,m^{-2}}\)) &
    Mean (\(\mathrm{mJ\,m^{-2}}\)) \\
    \midrule

    \multirow{3}{*}{100[010]}
    & 1 & \(926.1\pm1.9\) & \(112\pm168\)  & \(56\pm168\)  & 86.8 & \multirow{3}{*}{86.1} \\
    & 2 & \(925.9\pm1.9\) & \(155\pm172\)  & \(131\pm171\) & 87.0 & \\
    & 3 & \(926.0\pm2.0\) & \(-29\pm172\)  & \(18\pm173\)  & 84.5 & \\
    \hline

    \multirow{3}{*}{110[001]}
    & 1 & \(926.2\pm2.0\) & \(-133\pm178\) & \(-261\pm164\) & 87.8 & \multirow{3}{*}{89.0} \\
    & 2 & \(926.6\pm2.1\) & \(90\pm183\)   & \(91\pm171\)   & 90.8 & \\
    & 3 & \(926.1\pm2.0\) & \(295\pm178\)  & \(400\pm170\)  & 88.4 & \\
    \hline

    \multirow{3}{*}{110[1\(\bar{1}\)0]}
    & 1 & \(925.6\pm1.9\) & \(-41\pm166\)  & \(168\pm169\)  & 133.8 & \multirow{3}{*}{133.7} \\
    & 2 & \(925.8\pm2.0\) & \(-242\pm160\) & \(-146\pm165\) & 133.0 & \\
    & 3 & \(925.7\pm2.0\) & \(399\pm166\)  & \(167\pm167\)  & 134.2 & \\
    \hline

    \multirow{3}{*}{110[1\(\bar{1}\)2]}
    & 1 & \(926.9\pm2.2\) & \(257\pm185\) & \(199\pm181\) & 98.1  & \multirow{3}{*}{98.9} \\
    & 2 & \(926.7\pm2.1\) & \(72\pm190\)  & \(18\pm182\)  & 100.7 & \\
    & 3 & \(926.2\pm2.2\) & \(455\pm179\) & \(296\pm183\) & 97.9  & \\

    \bottomrule
\end{tabular}
\end{adjustbox}
\end{table}

To propagate replica variability into the anisotropy parameters, one stiffness value is selected from each of the three minimal orientations, \(100[010]\), \(110[001]\), and \(110[1\bar{1}0]\), with thickness of ~24\AA{}. Since three independent replicas are available for each orientation, this generates \(3^3=27\) possible anisotropy fits. These combinations should not be interpreted as 27 fully independent simulations, but rather as a propagation of the replica-to-replica variability in the orientation-resolved stiffnesses into \(\gamma_0\), \(\varepsilon_1\), and \(\varepsilon_2\).

\begin{figure}[t!]
    \centering
    \includegraphics[width=0.95\textwidth]{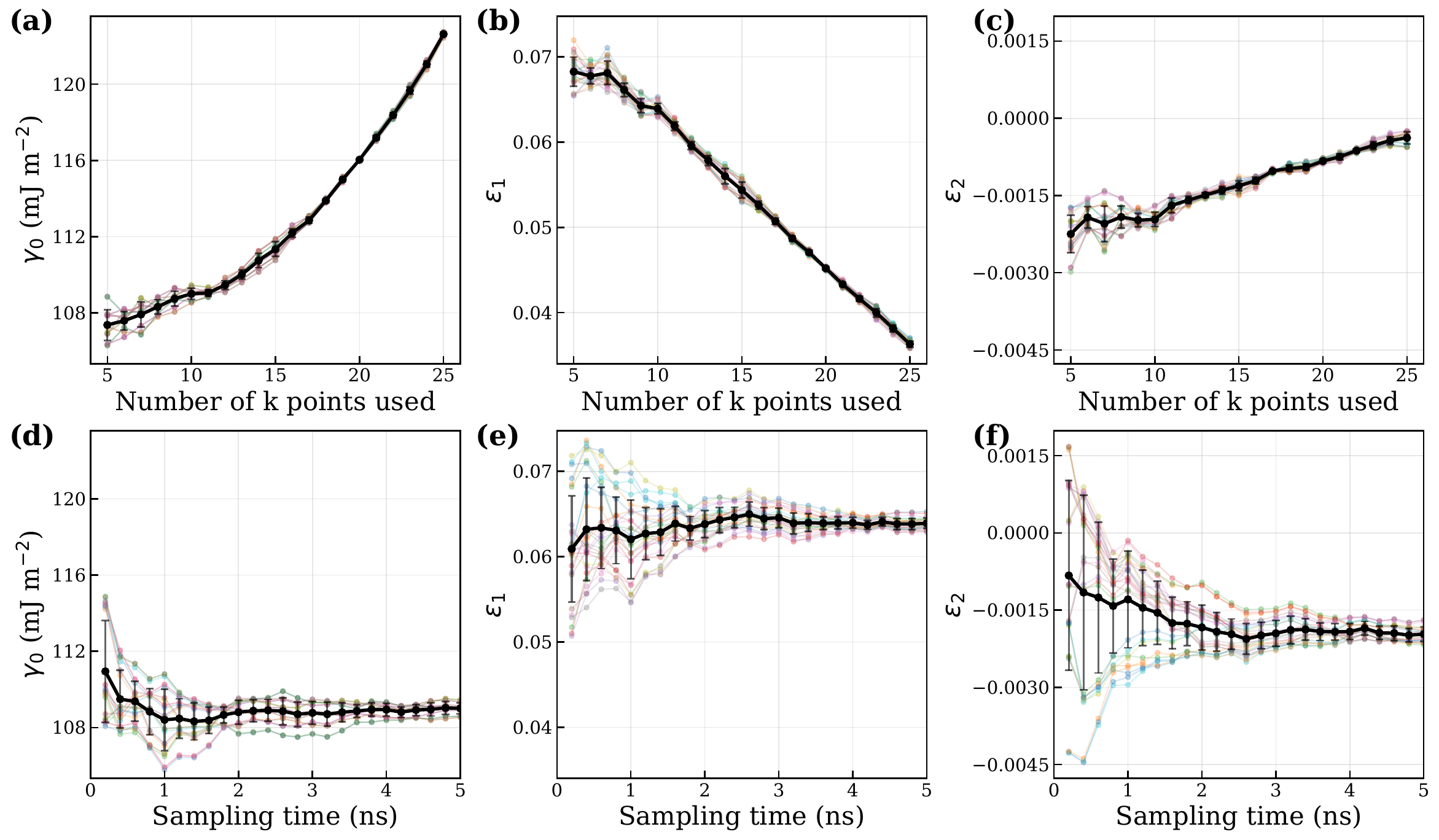}
    \caption{
    Replica and sampling-time uncertainty in fitted anisotropy parameters. (a--c) Effect of the number of \(k\)-points included in the stiffness fit on \(\gamma_0\), \(\varepsilon_1\), and \(\varepsilon_2\). Three independent replicas are performed for each of the \(100[010]\), \(110[001]\), and \(110[1\bar{1}0]\) interfaces; their stiffnesses are combined to generate 27 possible anisotropy fits. Colored thin curves denote individual replica combinations, black symbols denote the mean, and error bars denote the standard deviation over the combinations. (d--f) Effect of production sampling duration on the same fitted quantities using the selected fitting window. The large fluctuations at short sampling times reflect incomplete averaging of slowly relaxing capillary modes, whereas the residual spread at long sampling times reflects trajectory-to-trajectory variability associated with independent initial conditions and residual lateral stresses.
    }
    \label{fig:replica_sampling_uncertainty}
\end{figure}

Fig.~\ref{fig:replica_sampling_uncertainty}(a--c) shows the anisotropy fitting results obtained using different numbers of \(k\)-points in the stiffness fits. 
The black curves and error bars represent the mean values and standard deviations obtained by propagating the replica-to-replica variability of the orientation-resolved stiffnesses into \(\gamma_0\), \(\varepsilon_1\), and \(\varepsilon_2\). 
This uncertainty propagation shows that relatively small variations in individual stiffnesses can lead to a noticeable spread in the fitted anisotropy parameters. 
The effect is especially pronounced for \(\varepsilon_2\), which is obtained from differences between stiffnesses of similar magnitude and is therefore more sensitive to small systematic or statistical shifts.

As more \(k\)-points are included in the stiffness fits, the propagated standard deviation appears to decrease. 
This apparent reduction mainly reflects the fact that adding more fitted points reduces the influence of point-to-point variations on the regression. 
It should therefore not be interpreted by itself as convergence of the CFM result. 
Indeed, both \(\varepsilon_1\) and \(\varepsilon_2\) change systematically when higher-\(k\) modes are included, indicating that the fitted anisotropy parameters are still affected by the choice of fitting window. 
By contrast, when approximately 7--10 \(k\)-points are used, \(\gamma_0\), \(\varepsilon_1\), and \(\varepsilon_2\) fluctuate around relatively stable values, although the propagated uncertainty is larger in this conservative low-\(k\) range. 
Because these low-\(k\) modes are more consistent with the continuum assumptions of CFM and have passed the relaxation-time screening discussed above, this region is taken as the converged fitting window for the present analysis.

The spread among single-replica estimates illustrates the uncertainty that would be missed if only one trajectory were used for each orientation. 
Even after applying relaxation-time screening, conservative \(k\)-window selection, and consistent LOP-based SLI construction, replica selection can still change \(\varepsilon_1\) by approximately \(0.007\) and \(\varepsilon_2\) by approximately \(0.001\). 
These variations are comparable to subtle physical trends reported in previous CFM studies, such as the changes caused by composition and temperature differences in Al--Cu SLI~\cite{swamy2023atomistic}. Such uncertainty can propagate into phase-field simulations that use CFM-derived anisotropy parameters as input, potentially affecting predictions of dendrite growth direction and orientation selection, especially near regimes where competing orientations have similar stability. 

The same production trajectories are also analyzed using progressively longer time windows to separate finite-time convergence from residual trajectory-to-trajectory variability. As shown in Fig.~\ref{fig:replica_sampling_uncertainty}(d--f), the fitted anisotropy parameters fluctuate strongly when the sampling duration is shorter than approximately \(2~\mathrm{ns}\). For sampling durations longer than approximately \(2~\mathrm{ns}\), the mean values enter a more stable regime. The remaining spread at long sampling times reflects residual trajectory-to-trajectory variability associated with independent initial configurations and residual lateral stress states, rather than finite-time sampling alone.

\subsection*{Convergence of interfacial parameters}
\label{sec:convergence}

We next examine whether the three fitted interfacial parameters, $\gamma_0$, $\varepsilon_1$, and $\varepsilon_2$, converge with model thickness and interface representation when replica-based uncertainty is taken into account. 
For the ribbon-model analysis, we use the combination \(100[010]+110[001]+110[1\bar{1}0]\), which provides a reasonably well-conditioned minimal basis while avoiding orientations that exhibit
slower thickness convergence. The stiffness relations and conditioning of alternative orientation combinations are summarized in Section S2 of the Supplementary Information.

\begin{figure}
    \centering
    \includegraphics[width=1\linewidth]{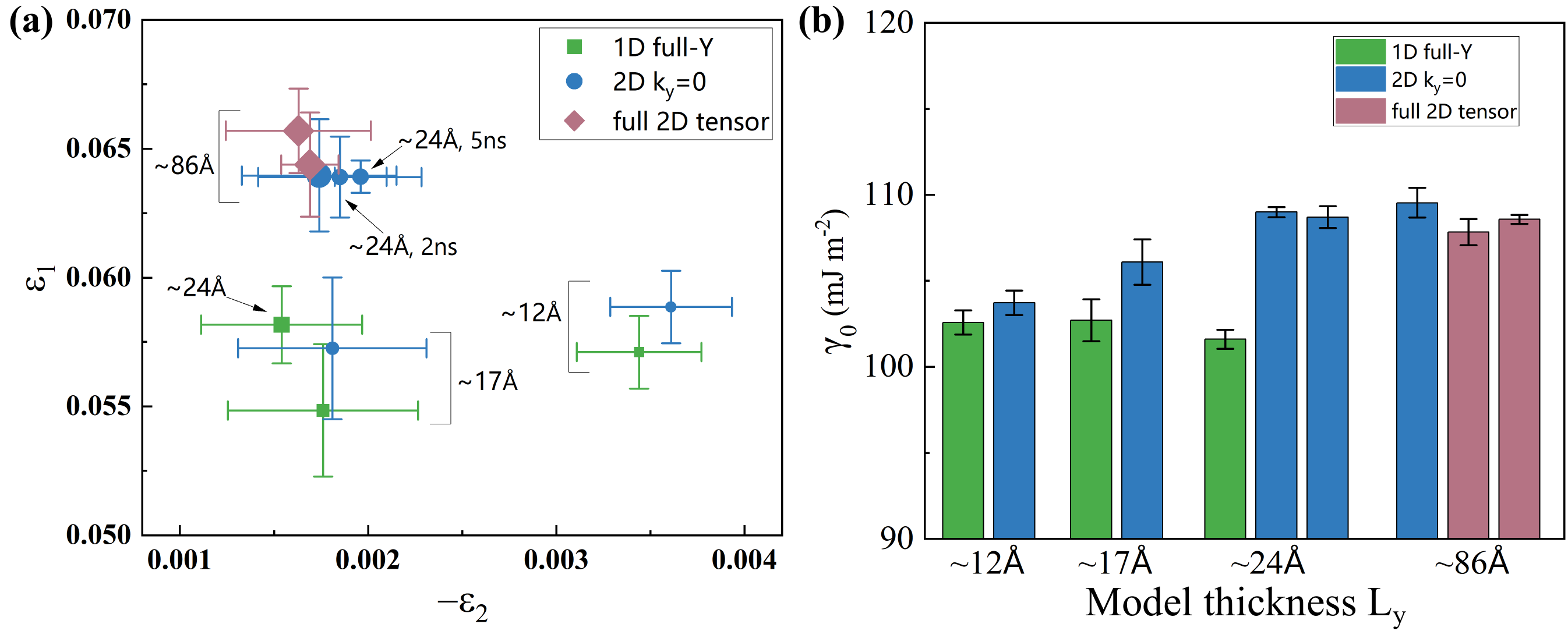}
    \caption{
Comparison of the interfacial parameters obtained using different model thicknesses and interface representations.
(a) Anisotropy parameters $\varepsilon_1$ and $\varepsilon_2$, and
(b) orientation-averaged interfacial free energy $\gamma_0$.
Green, blue, and purple symbols denote results obtained using the 1D full-$Y$, 2D $k_y=0$, and full 2D tensor representations, respectively.
Error bars indicate the standard deviation arising from the replica-based uncertainty propagation described in the preceding section.
For the 1D full-$Y$ and 2D $k_y=0$ analyses, each parameter set is obtained from all $3^3=27$ combinations of the three independent replicas for the three primary orientations, whereas $3^2=9$ combinations are used for each full 2D tensor result.
In panel (a), symbol size represents the model thickness.
For the $\sim24$~\AA{} models, results from both 2 and 5~ns production trajectories are shown; their difference, most visibly in $\varepsilon_2$, is substantially smaller than the replica-based uncertainty.
Two full 2D tensor estimates are shown because two independent 110-type tensor representations, with $L_x$ along $\langle110\rangle$ and $\langle001\rangle$, were available for the anisotropy fit.
The $\sim24$~\AA{} 2D $k_y=0$ results converge closely to those obtained from the $\sim86$~\AA{} full 2D tensor models, both for the anisotropy parameters in panel (a) and for $\gamma_0$ in panel (b).
}
    \label{fig:Three_parameters_convergence}
\end{figure}

The results are summarized in Fig.~\ref{fig:Three_parameters_convergence}. 
Panel (a) shows the fitted anisotropy parameters $\varepsilon_1$ and $\varepsilon_2$, while panel (b) shows the orientation-averaged interfacial free energy $\gamma_0$. 
Green, blue, and purple symbols denote the 1D full-$Y$, 2D $k_y=0$, and full 2D tensor representations, respectively. 
The symbol size in panel (a) indicates the model thickness. 
For the 1D full-$Y$ and 2D $k_y=0$ analyses, each point represents the mean over all $3^3=27$ combinations of the three replicas for the three primary orientations, with the error bars showing the corresponding standard deviation. 
For the full 2D tensor analysis, each point is obtained from $3^2=9$ replica combinations. 
Two full-tensor estimates are shown because two independent 110-type tensor models, with the in-plane fluctuation direction along $\langle110\rangle$ and $\langle001\rangle$, respectively, can be combined with the 100-type tensor result.

A clear distinction emerges between the interface representations. 
The parameters obtained using the 1D full-$Y$ treatment remain strongly dependent on model thickness, and the discrepancy between the 1D full-$Y$ and 2D $k_y=0$ results becomes increasingly pronounced as $L_y$ increases. 
This behavior is consistent with the stiffness-level comparison in Fig.~\ref{fig:thickness_convergence}, where the 1D full-$Y$ representation does not show systematic convergence toward the thick-model results.
In contrast, the parameters obtained using the 2D $k_y=0$ representation converge rapidly with increasing thickness. 
For the $\sim24$~\AA{} ribbon models, both anisotropy parameters already lie close to the values obtained from the $\sim86$~\AA{} models. 
The two independent full 2D tensor estimates from the thick models are also mutually consistent within their propagated uncertainties and cluster closely with the 2D $k_y=0$ results from both the $\sim24$ and $\sim86$~\AA{} models. 
Based on the stiffness-level consistency established above, the thick-model full 2D tensor results are therefore taken as the reference estimates in the following discussion.
The $\sim24$~\AA{} 2D $k_y=0$ calculations performed using 2 and 5~ns production trajectories also give nearly identical fitted parameters. 
The largest visible difference occurs for $\varepsilon_2$, but remains much smaller than the replica-based uncertainty. 
This provides an independent confirmation that the 2~ns production duration adopted in the convergence tests is sufficient for the present ribbon models. The same convergence trend is observed for $\gamma_0$ in Fig.~\ref{fig:Three_parameters_convergence}(b). 
The 1D full-$Y$ estimates retain a pronounced thickness dependence, whereas the 2D $k_y=0$ value obtained from the $\sim24$~\AA{} ribbon models is already close to those obtained from the much thicker models and from the full 2D tensor analysis.

Taken together, these results show that a ribbon geometry can reproduce the interfacial parameters of the thick 2D reference provided that the transverse dimension is sufficiently converged and the interface is first resolved in two dimensions before restricting the analysis to the $k_y=0$ modes.

\section*{Discussion}
\label{sec:discussion}

Our results show that the reliability of CFM calculations should be assessed through a set of diagnostic checks rather than through regression quality alone. The main methodological implications are discussed below.

Let us first consider the selection of the fitting \(k\)-window.
Previous CFM studies have commonly used the apparent \(k^{-2}\) scaling of the fluctuation spectrum, often inspected on log--log plots of \(\langle |A(k)|^2\rangle\)~\cite{hoyt2001method,asta2002calculation,swamy2023atomistic,urashima2026capillary}, as a primary criterion for selecting the fitting range. 
This criterion is necessary but not sufficient. 
Logarithmic representations can visually compress systematic deviations, and modes that follow an apparent power law may still yield different stiffnesses when different \(k\)-windows are used. 
The sensitivity of CFM results to fitting-range selection has been recognized previously. 
For example, Rozas \textit{et al.} examined different \(k\)-ranges and used the small-\(k\) limiting behavior as a reference~\cite{rozas2023interfacial}. 
However, the smallest accessible modes are not automatically the most reliable, because their long relaxation times can lead to insufficient temporal sampling within finite MD trajectories. 
Conversely, extending the fit to larger \(k\) can improve the number of fitted modes but may introduce bias from finite interface width, atomistic noise, and interface-construction artifacts.

This trade-off also helps explain why different fitting windows have been adopted in previous CFM studies. 
Using different \(k\)-windows for different orientations~\cite{asadi2015two,azizi2022interactive} can be practically convenient, especially when the accessible wave vectors and fluctuation spectra differ between simulation cells. 
However, without common diagnostics, such choices introduce additional subjectivity into anisotropy fitting. 
In the present Al demonstration, the validated fitting window, \(0.005 < k^2 < 0.03~\mathrm{\AA^{-2}}\), is narrower than the upper \(k^2\) ranges used in some previous Al CFM studies, such as \(k^2 \lesssim 0.1~\mathrm{\AA^{-2}}\)~\cite{asadi2015two} and \(k^2 \lesssim 0.08~\mathrm{\AA^{-2}}\)~\cite{urashima2026capillary}. 
These numerical ranges are not directly transferable across studies because they depend on simulation geometry, interface definition, sampling protocol, and interatomic potential. 
Nevertheless, the comparison highlights that the high-\(k\) cutoff is a nontrivial methodological choice rather than a purely technical fitting parameter.

The practical implication is that the fitting window should be selected as the overlap between two admissible ranges: modes that are temporally resolved and modes that remain stable with respect to interface construction. 
In this workflow, relaxation-time analysis identifies poorly sampled low-\(k\) modes, whereas \(k\)-window and coarse-graining sensitivity tests detect high-\(k\) bias. 
A sufficiently dense set of accessible wave vectors further improves this diagnosis, because deviations from linearity and fitting-window dependence can be assessed more clearly than in fits based on only a few modes.

A second implication is that uncertainty should be propagated from orientation-resolved stiffnesses to the final anisotropy parameters. 
The parameters \(\varepsilon_1\) and \(\varepsilon_2\) are obtained by combining stiffnesses from different crystallographic orientations, and therefore small shifts in individual stiffnesses can be amplified during anisotropy fitting. 
This amplification is especially important for parameters determined from differences between stiffnesses of comparable magnitude.

Statistical uncertainty in the CFM has often been estimated directly from the fluctuation spectrum. 
For example, the standard uncertainty of each spectral amplitude can be estimated as~\cite{asta2002calculation,hoyt2003atomistic}
\begin{equation}
\sigma_{\langle |A(k)|^2\rangle}
=
\langle |A(k)|^2\rangle
\sqrt{\frac{2\tau(k)}{t_{\mathrm{run}}}},
\end{equation}
where \(t_{\mathrm{run}}\) is the sampling time and \(\tau(k)\) is the relaxation time of the corresponding Fourier mode. 
This expression accounts for the reduced number of statistically independent samples associated with slowly relaxing low-\(k\) modes. 
Other studies have estimated uncertainties by averaging the fluctuation amplitude at each \(k\) and reporting the corresponding standard deviation~\cite{swamy2023atomistic,morris2002anisotropic}. 
Such approaches are useful for quantifying finite-sampling uncertainty within a given trajectory, but they do not by themselves capture trajectory-to-trajectory variability or systematic variability introduced during model preparation.

Independent replicas therefore provide complementary information. 
They sample variations associated with initial configurations, liquid-slab placement, velocity seeds, coexistence conditions, and residual lateral stresses, which may not be fully represented by the uncertainty of a single fluctuation spectrum. 
Extending one trajectory can reduce finite-time noise, but it does not necessarily test whether the resulting stiffnesses and anisotropy parameters are reproducible with respect to independent realizations of the two-phase system. 
The close agreement between the parameters obtained from 2 and 5~ns trajectories for the \(\sim24\)~\AA{} ribbon models further indicates that, once the relevant capillary modes are adequately sampled, replica-to-replica variability can become more important than simply extending an individual trajectory.

Third, the comparison between ribbon and thick 2D models is influenced by two distinct but coupled sources of thickness dependence: one arising from the interface representation used in post-processing, and the other from the finite dimensions of the simulated system itself. 
These two effects need to be separated before the convergence of ribbon-model CFM calculations can be assessed.
The first effect originates from the treatment of the transverse direction during interface construction. 
The conventional 1D full-\(Y\) approach has been widely used for thin ribbon geometries, but its sensitivity to model thickness has also been noted previously~\cite{ambler2017solid}. 
In this representation, information along the transverse direction is discarded before the interface is identified. 
Although this approximation can be reasonable for very thin ribbons, it becomes increasingly restrictive as transverse variations of the interface develop with increasing \(L_y\), leading to an apparent thickness dependence that originates from the analysis procedure rather than necessarily from the simulated system itself. 
The 2D \(k_y=0\) representation provides a more physically consistent treatment for ribbon models: the interface is first resolved as \(h(x,y)\), preserving transverse information during interface construction, and only the \(k_y=0\) modes are retained for the stiffness analysis. 
At the other limit, the full 2D tensor approach makes the most complete use of the fluctuation spectrum, but requires substantially thicker simulation cells so that nonzero \(k_y\) modes fall within the accepted low-\(k\) range. 
The 2D \(k_y=0\) representation therefore provides an intermediate strategy that retains the computational advantage of ribbon geometries while avoiding the analysis-induced contribution to the thickness dependence.

Once this post-processing contribution is separated, the 2D \(k_y=0\) results reveal a second thickness dependence associated with the simulation geometry itself. 
The calculated stiffnesses continue to change with increasing \(L_y\) even after the model thickness exceeds twice the interaction cutoff, and eventually converge toward the thick-model values. 
For the 100[010], 110[001], and 110[1\(\bar{1}\)0] orientations, the \(\sim24\)~\AA{} ribbon models reproduce the corresponding \(\sim86\)~\AA{} stiffnesses to within approximately \(1\%\). 
The full 2D tensor results obtained from the thick models are also consistent with both the thick-model \(k_y=0\) results and the converged ribbon values, supporting their use as reference estimates. 
Thus, \(L_y>2r_{\mathrm{cut}}\) should be regarded only as a minimum geometric requirement for avoiding direct periodic self-interactions, rather than as a criterion for CFM thickness convergence.

The convergence thickness is also orientation dependent. 
Most notably, the 110[1\(\bar{1}\)2] stiffness remains measurably different from the thick-model value even when \(L_y\) is increased to approximately 35~\AA{}. 
This observation has an important implication for the use of redundant crystallographic orientations. 
Although stiffnesses from different fluctuation directions of the same crystallographic family satisfy exact linear relations within the cubic-harmonic description, these relations need not be recovered by finite simulation cells if the corresponding orientations have not individually reached their thickness-converged regimes. 
Agreement among several similarly thin models therefore does not demonstrate convergence by itself.

The solid--liquid coexistence temperature provides a useful indicator of this finite-size effect. 
In the thermodynamic limit, the equilibrium melting temperature at fixed pressure and composition should be independent of interface orientation, whereas finite simulation cells can exhibit orientation-dependent apparent coexistence temperatures. 
Such differences have also been reported in previous CFM studies. 
In the present calculations, the thinnest ribbon models show a broad distribution of equilibrium temperatures among orientations, whereas this variation decreases systematically with increasing \(L_y\). The reduction in the spread of the mean coexistence temperatures cannot be explained simply by the increase in system size and the corresponding reduction in instantaneous temperature fluctuations. 
The latter follow the expected statistical scaling with system size, whereas the systematic shift of the mean temperature reflects a distinct finite-size modification of the two-phase equilibrium state. 
We therefore propose the coexistence temperature as a practical diagnostic for selecting the transverse dimension of ribbon models. 
As a rule of thumb, the ribbon thickness should (i) exceed \(2r_{\mathrm{cut}}\), (ii) yield mutually consistent coexistence temperatures among the investigated orientations, and (iii) reproduce the coexistence temperature of a sufficiently thick reference system. 
For the present Al system, these conditions are satisfied at approximately 24~\AA{} for the three orientations used in the primary anisotropy fit, consistent with their stiffness convergence.

Benefiting from the preceding diagnostics, the final anisotropy parameters retain low standard deviations even after replica-to-replica uncertainty is included. 
In this work, the maximum standard deviations among the tested orientation combinations are \(0.0015\) for \(\varepsilon_1\) and \(0.0004\) for \(\varepsilon_2\) (24~\AA{} models, 2D \(k_y=0\) method). 
These values are lower than the uncertainty levels reported in several previous CFM studies of Al and Al-alloy SLIs, where standard deviations of \(0.006\)--\(0.02\) for \(\varepsilon_1\) and \(0.0015\)--\(0.003\) for \(\varepsilon_2\) have been reported~\cite{wang2020controlling,swamy2023atomistic}. 
Although these values are not directly interchangeable, the comparison suggests that the proposed diagnostics-driven workflow can reduce avoidable variability while still providing a realistic estimate of trajectory-to-trajectory uncertainty.

The resulting parameters are also consistent with the broad range of values reported previously for Al. The present work gives \(\gamma_0=108.7~\mathrm{mJ\,m^{-2}}\), \(\varepsilon_1=0.0639\), and \(\varepsilon_2=-0.00185\).
Experimental estimates of the orientation-averaged SLI free energy span approximately \(93\)--\(158~\mathrm{mJ\,m^{-2}}\), while atomistic calculations using different interatomic potentials and methodologies likewise show substantial scatter, with representative values ranging from approximately \(98\) to \(173~\mathrm{mJ\,m^{-2}}\)~\cite{yan2020solid,fan2026solid}.
Zhong \emph{et al.} obtained \(\varepsilon_1\simeq0.060\) and \(\varepsilon_2\simeq-0.0036\) for pure Al using the same Al--Si AEAM potential~\cite{zhong2025quantification}.
Importantly, reliable experimental benchmarks for the anisotropy of metallic SLIs are still unavailable, and even the reported orientation-averaged interfacial free energies remain strongly method dependent~\cite{fan2026solid}.
This substantial spread further highlights the value of a diagnostics-based CFM workflow that explicitly identifies and controls methodological sources of variability, rather than relying solely on apparent statistical precision.

The practical outcome of this work is not a single universal set of numerical parameters, but a set of diagnostics that should accompany reported CFM-derived values. At minimum, CFM studies should report the selected \(k\)-window, relaxation times of retained modes, production duration, output interval, descriptor and coarse-graining parameters, the method used to represent the interface, replica variability, residual lateral stresses, model dimensions, interaction cutoff, coexistence behavior, and thickness-convergence tests.
When thick 2D calculations are feasible, comparison between \(k_y=0\) and full tensor results provides an additional reference-level validation.
Reporting these diagnostics would make comparisons between different potentials, alloy systems, and simulation protocols more meaningful.

These considerations are particularly relevant for MLIP-based CFM simulations. Although MLIPs can provide improved accuracy and extend the CFM to more complex systems, they are typically about one order of magnitude slower than empirical potentials~\cite{shuang2025universal}. This is important because the CFM requires both large simulation cells and long sampling trajectories. A converged ribbon geometry can therefore provide a substantial efficiency advantage. 
Compared with a nearly square 2D interface capable of accessing a similar low-$k$ range in both directions, the workload is reduced by approximately a factor of 25. In addition, because the CFM requires long equilibrium trajectories near the melting point, long-time energy conservation and coexistence stability should be verified carefully~\cite{leimeroth2025machine}.

To facilitate practical implementation, an open-source Python package is provided for interface construction, fluctuation analysis, relaxation-time fitting, anisotropy fitting, uncertainty propagation, and diagnostic plotting. The workflow and tools developed here are expected to improve the reproducibility and comparability of CFM calculations, particularly for chemically complex interfaces and MLIP-based simulations.

\section*{Methods}\label{sec:methods}


\subsection*{Interatomic potential}
The proposed workflow is demonstrated using pure Al described by the angular-dependent embedded-atom method (AEAM) potential developed by Saidi \textit{et al.}~\cite{saidi2014angular}. This potential has recently been used in capillary-fluctuation studies of Al-based SLIs~\cite{zhong2025quantification}. The Al-Al pair cutoff distance is 6.5~\AA{}. Since the performance of the AEAM potential has already been benchmarked in previous studies, the present demonstration does not include additional validation. Instead, the focus is on the methodological sensitivity and uncertainty propagation of the CFM analysis. The initial melting temperature is taken as \(T_m=927~\mathrm{K}\), and all MD simulations are performed using LAMMPS~\cite{LAMMPS} with a time step of \(1~\mathrm{fs}\).

\subsection*{Simulation-cell geometry and dimensions}
The simulation cells contain two solid--liquid interfaces with their mean normals aligned along the \(z\)-direction. The \(x\)- and \(y\)-directions span the interfacial plane, with \(x\) chosen as the primary fluctuation direction and \(y\) as the transverse direction. The instantaneous SLI can generally be represented as a two-dimensional height field \(h(x,y)\). Depending on the model thickness and analysis method, this field is either reduced to its \(k_y=0\) component or analyzed using the full two-dimensional fluctuation spectrum, as described below.

The interfacial length \(L_x\) determines the smallest accessible wave vector and the spacing between neighboring modes, \(\Delta k=2\pi/L_x\). Increasing \(L_x\) therefore provides a denser set of modes within a given fitting window, although at increased computational cost and with slower relaxation of the longest-wavelength modes.

The transverse dimension \(L_y\) controls both finite-size effects and the accessible fluctuations along the second in-plane direction. The condition \(L_y>2r_{\rm cut}\), where \(r_{\rm cut}=6.5\)~\AA{} for the AEAM potential used here, is treated only as a minimum geometric requirement to avoid direct periodic self-interactions. As demonstrated in the Results, thickness convergence must be assessed explicitly rather than inferred from the interaction cutoff alone.

The interface-normal dimension \(L_z\) is chosen sufficiently large to maintain well-separated solid--liquid interfaces and extended solid and liquid regions between them.

In the present calculations, \(L_x\) was kept close to \(585\)~\AA{} and \(L_z\) close to \(180\)~\AA{} across the investigated orientations, while \(L_y\) was systematically varied from approximately \(12\) to \(86\)~\AA{}. The thinner cells constitute the ribbon-model thickness series, whereas the thickest cells serve as two-dimensional reference models. The detailed model set is summarized in Fig.~\ref{fig:model_thickness_design}.

Following the naming convention of Ref.~\cite{hoyt2001method}, each model orientation is labeled by two Miller indices of the solid phase, for example, 100[010], where the first index denotes the interface normal (\(z\)) and the second denotes the designated in-plane fluctuation direction (\(x\)).

\subsection*{Orientation selection and anisotropy fitting}
Because Eq.~\eqref{Eq:cubic_harmonic_expansion} contains three unknown parameters, \(\gamma_0\), \(\varepsilon_1\), and \(\varepsilon_2\), at least three independent stiffness relations are required. These may be obtained from three ribbon models or, when multiple in-plane stiffness components are available, from two sufficiently thick 2D interface models. In practice, the commonly used set $100[010]$, $110[001]$, and $110[1\bar{1}0]$ provides a convenient minimal basis to determine the leading anisotropy parameters.
Other orientations, such as \(110[1\bar{1}2]\) and \(111[1\bar{2}1]\), have also been used in the literature~\cite{yan2020solid,yin2024far}. Higher-order expansions can be introduced when a larger set of independent stiffness measurements is available (for example, an eighth-order expansion was presented in \cite{rozas2023interfacial}), but the two-parameter form remains the standard choice for most practical CFM studies. $111$-oriented SLI may become faceted rather than thermally rough~\cite{frolov2012step}, in which case the interfacial thermodynamics is governed by step properties and the conventional CFM framework is no longer directly applicable.

In the present work, the set \(100[010]\), \(110[001]\), and \(110[1\bar{1}0]\) provides a well-conditioned minimal basis, whereas \(110[1\bar{1}2]\) is used primarily to assess orientation-dependent thickness convergence. The stiffness relations and condition numbers for the tested orientation combinations are summarized in the Supplementary Information S2.

\subsection*{Ensemble, coexistence condition, and fixed wave vectors}
The validity of the CFM relies on proper sampling of thermal equilibrium fluctuations at solid--liquid coexistence. In principle, different ensembles, such as NVE~\cite{hoyt2001method}, NPH~\cite{asadi2015two}, or NPT~\cite{swamy2023atomistic,dolce2023computing}, can be employed, provided that coexistence is maintained and that the thermostat or barostat does not artificially distort the interfacial fluctuations of interest. For static stiffness calculations, the primary requirement is correct equilibrium sampling of the fluctuation spectrum. For relaxation-time analysis, however, the choice of thermostat may additionally affect the temporal correlation of capillary modes and should therefore be treated with caution. Two points require particular attention.

First, the determination of the coexistence temperature in finite slab models is nontrivial. In finite MD systems, the apparent melting temperature \(T_m\) is often observed as a distribution rather than a single value~\cite{zhu2021fully}, and CFM models with different crystal orientations may exhibit slight variations in their stable coexistence temperatures~\cite{haapalehto2022atomistic,yan2020solid,yin2024far}. Therefore, applying a thermostat at a bulk-determined \(T_m\) during production sampling can introduce a non-equilibrium bias, for example by causing slow melting or solidification of the two-phase system. Some studies therefore determine the coexistence temperature separately for each slab geometry. An alternative strategy is to use an isenthalpic--isobaric ensemble (NPH), which allows the system temperature to adjust naturally while maintaining phase coexistence~\cite{asadi2015two}.

Second, the accessible wave vectors must remain fixed during production sampling. According to Eq.~\eqref{Eq:Lx_k_selection}, the allowed wave vectors are determined by the interfacial length \(L_x\). In the quasi-two-dimensional geometry, the thin dimension \(L_y\) also enters the fluctuation spectrum through the interfacial area factor \(L_xL_y\). Therefore, the lateral dimensions used to define the capillary modes and the interfacial area should remain fixed during sampling. By contrast, the interface-normal dimension \(L_z\) may be allowed to relax in order to maintain the desired normal pressure and stable coexistence. This constraint, however, can leave a small residual lateral stress in the fixed directions, which should be treated as a possible source of trajectory-to-trajectory variability and systematic uncertainty. Its influence is evaluated in the Results, where corresponding workflow recommendations are discussed.

To build a two-phase solid--liquid configuration from an FCC crystal, atoms in the central region of the simulation box, approximately \(0.25<z/L_z<0.75\), are heated to \(T_m+500~\mathrm{K}\), while the remaining atoms are frozen. This melting stage is performed for \(100~\mathrm{ps}\) using an NVE integrator with a Langevin thermostat. The melted region is then gradually cooled back to \(T_m\).

\begin{figure}[!t]
    \centering
    \includegraphics[width=0.75\linewidth]{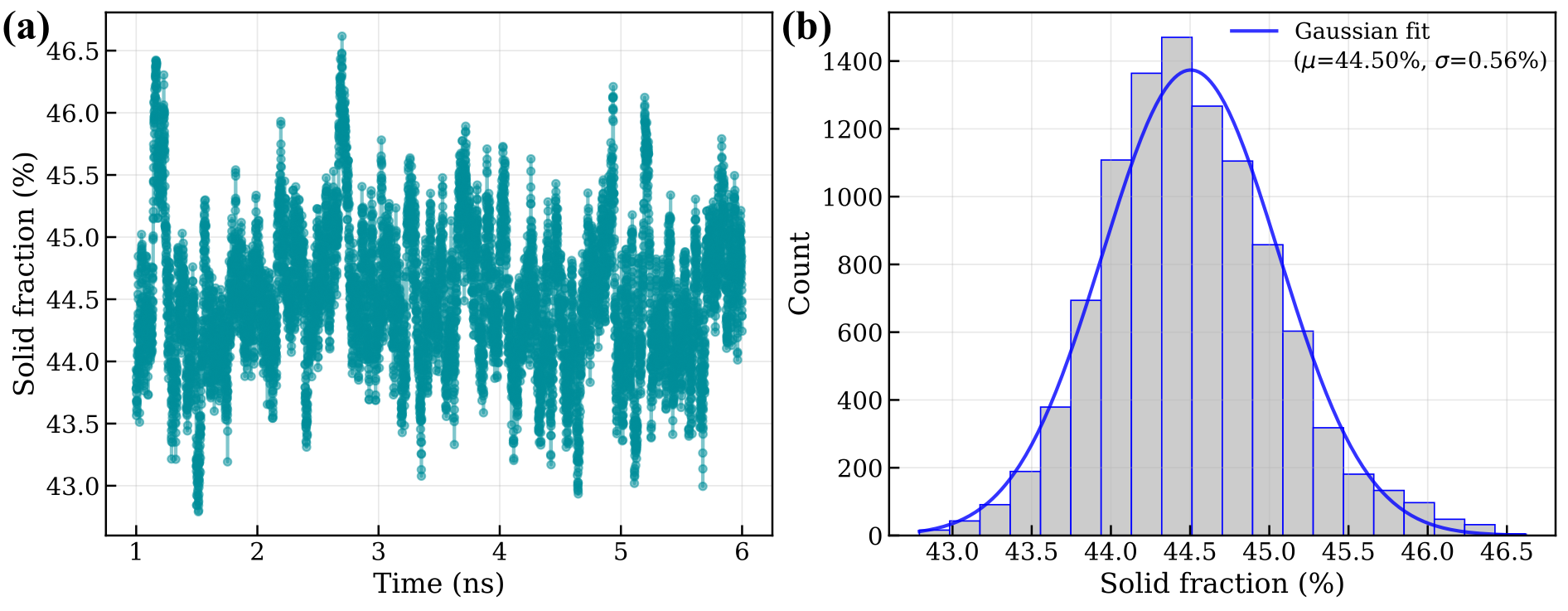}
\caption{
Time evolution and distribution of the solid-phase volume during equilibrium CFM sampling.
The solid fraction (a) remains statistically stationary over the production trajectory and (b) exhibits a narrow, approximately normal distribution, indicating that the system remains close to solid--liquid coexistence without systematic melting or solidification.
This provides a practical sanity check for the two-phase simulation before applying capillary fluctuation analysis.
The solid-volume distribution is not used directly to determine the interfacial stiffness, but serves as a diagnostic for the stability of the coexistence state.
}
\label{fig:S_solid_fraction}
\end{figure}

After the initial two-phase configuration is generated, the full system is equilibrated at \(T_m\) using an NPT ensemble for \(50~\mathrm{ps}\), allowing all box dimensions to relax. The system is then further equilibrated for \(1~\mathrm{ns}\) using an \(NP_zAH\)-type ensemble, where \(A=L_xL_y\) denotes the fixed interfacial area. Production sampling is subsequently carried out for \(2~\mathrm{ns}\), with atomic configurations saved every \(0.5~\mathrm{ps}\). A longer \(5~\mathrm{ns}\) production run was additionally performed for the nominal \(24~\mathrm{\AA}\) ribbon models to evaluate sampling-time convergence. During production, the solid fraction should be monitored for every trajectory as a practical coexistence diagnostic. 
A statistically stationary solid fraction, with a narrow, approximately normal distribution, indicates that the two-phase system remains close to coexistence without systematic melting or solidification. 
Fig.~\ref{fig:S_solid_fraction} shows a representative example for the 100[010] interface, where the solid fraction remains stable throughout the production run. 
This check should be performed for each independent simulation before applying capillary-fluctuation analysis.

\subsection*{Interface construction from atomistic configurations}
\label{sec:interface_construction}
In atomistic simulations, the SLI is inherently defined by discrete atomic configurations. However, the capillary fluctuation method assumes a continuous and well-defined interface. Therefore, it is necessary to construct a smooth and effectively sharp interface representation from atomistic data. This process involves three key steps: atom-wise structural identification, spatial coarse-graining, and extraction of the interface position.

\subsection*{Atom-wise descriptor}
At solid--liquid coexistence, thermal vibrations near the melting point make the distinction between solid-like and liquid-like atoms nontrivial. A robust atom-wise structural descriptor is therefore required before constructing a continuous interface field. Common atom-wise structural descriptors include common neighbour analysis (CNA)~\cite{faken1994systematic}, 
the centrosymmetry parameter (CSP)~\cite{kelchner1998dislocation}, \(q_6\)~\cite{steinhardt1983bond}, \(\bar{q}_6\)~\cite{lechner2008accurate}, PTM~\cite{larsen2016robust}, and LOP~\cite{hoyt2001method}.

When the crystallographic orientation of the solid phase is known, an orientation-specific LOP can be constructed by measuring the deviation of local neighbour vectors from their ideal lattice positions~\cite{hoyt2001method}. For an FCC crystal, for example, this descriptor can be written as
\begin{equation}
\phi_i = \frac{1}{12} \sum_{j=1}^{12} 
\left| \mathbf{r}_{ij} - \mathbf{r}_{ij}^{\mathrm{FCC}} \right|^2 ,
\end{equation}
where \(\mathbf{r}_{ij}\) is the vector from atom \(i\) to its \(j\)-th nearest neighbour, and \(\mathbf{r}_{ij}^{\mathrm{FCC}}\) is the corresponding ideal neighbour vector in the reference FCC lattice. Smaller values of \(\phi_i\) indicate a more solid-like local environment, whereas larger values correspond to stronger disorder.

Compared with more general structural classifiers such as PTM, an orientation-specific LOP directly uses the known crystallographic reference frame of the solid slab. This makes it particularly suitable for CFM simulations, where the solid orientation is prescribed and remains well defined during coexistence sampling. In contrast, more generic descriptors may introduce additional classification noise near the diffuse SLI.

Some studies further reduce thermal noise by temporally averaging atomic positions~\cite{brown2019solid} or quenching configurations to 0 K before structural analysis~\cite{mishin2014calculation}. Although such procedures can sharpen the apparent structural contrast, they may also suppress physically meaningful thermal fluctuations. Therefore, preprocessing steps that modify the instantaneous thermal configuration should be avoided in equilibrium capillary-fluctuation analysis.

\subsection*{Coarse graining}
\begin{figure}[t!]
    \centering
    \includegraphics[width=0.7\linewidth]{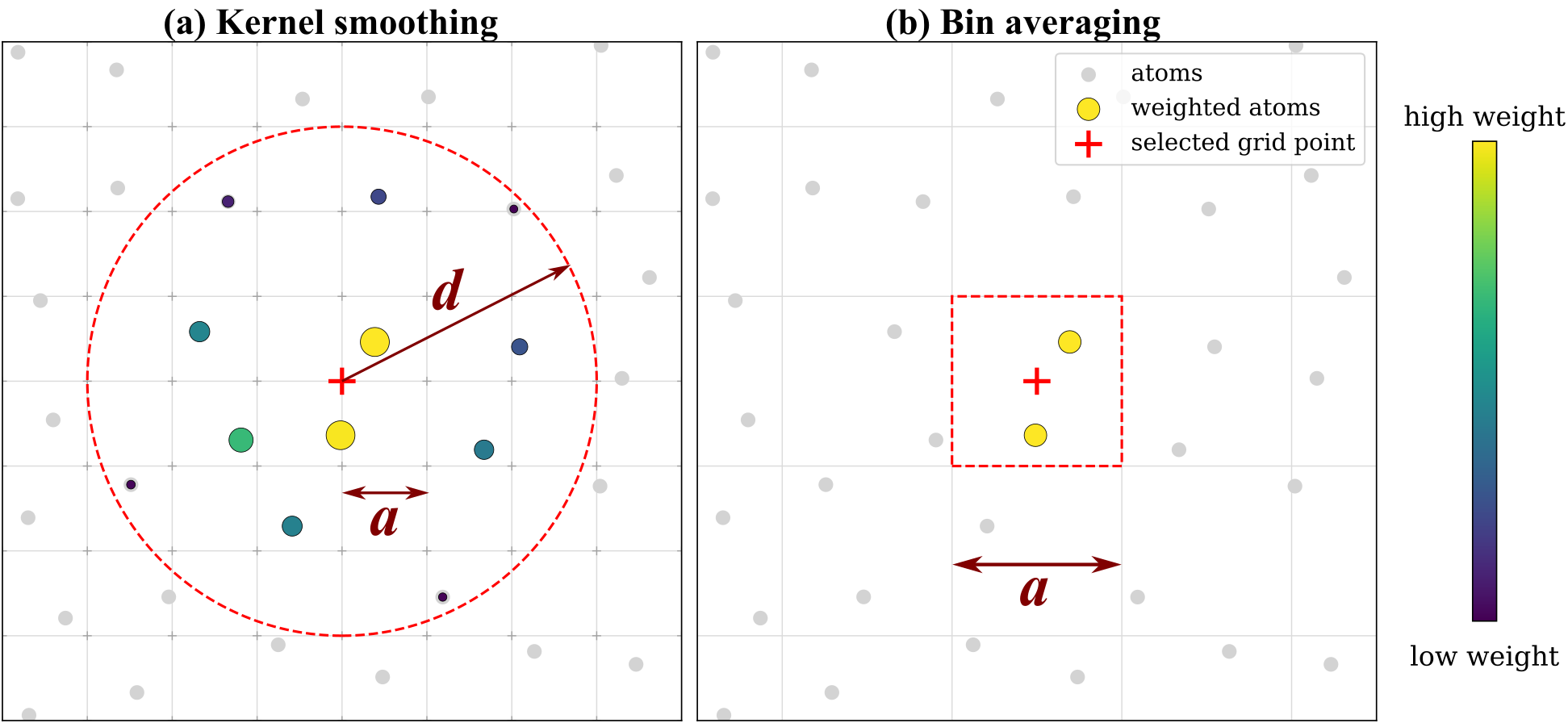}
    \caption{Coarse-graining strategies for constructing a continuous order-parameter field from atom-wise descriptors. 
    (a) In kernel smoothing, atoms within a smoothing radius \(d\) contribute with distance-dependent weights, while the grid spacing \(a\) controls only the spatial sampling of the coarse-grained field. 
    (b) In bin averaging, atoms inside each bin contribute with equal weight, so the bin size \(a\) simultaneously controls both the sampling resolution and the averaging volume.}
    \label{fig:kernel_vs_bin_averaging}
\end{figure}

After atom-wise descriptors are assigned, they must be mapped onto a continuous spatial field before the interface position can be extracted. This coarse-graining step is not without consequence: it introduces an effective spatial filter that mainly affects the short-wavelength part of the interface profile. Two commonly used approaches are bin averaging and kernel smoothing, as illustrated in Fig.~\ref{fig:kernel_vs_bin_averaging}.

In kernel smoothing, the field value at a grid point is computed as a weighted average over atoms within a smoothing radius \(d\)~\cite{davidchack2006anisotropic,asadi2015two}:
\begin{equation}
    w = \frac{\sum_i w_d(r_i)\,\phi_i}{\sum_i w_d(r_i)},
\end{equation}
with
\begin{equation}
    w_d(r_i) = \left[1 - \left(\frac{r_i}{d}\right)^2 \right]^2 ,
\end{equation}
where \(r_i\) is the distance between atom \(i\) and the grid point, and \(\phi_i\) is the atom-wise descriptor. This approach decouples the averaging length scale \(d\) from the grid spacing \(a\), allowing the field to be sampled finely while maintaining a controlled degree of smoothing.

In bin averaging, by contrast, descriptors are averaged over atoms inside discrete spatial bins. The bin size therefore controls both the spatial resolution and the number of atoms contributing to each average. Excessively small bins can lead to noisy fields because too few atoms are sampled, whereas excessively large bins can suppress physically meaningful short-wavelength fluctuations.

\subsection*{Interface representation}
Three approaches can be distinguished according to how fluctuations along the transverse direction \(y\) are treated. 
In the conventional \emph{1D full-\(Y\)} approach, commonly used for thin ribbon models, the transverse direction is ignored during interface construction and the interface is extracted directly as a one-dimensional profile \(h(x)\). 
The sensitivity of this treatment to the model thickness has been emphasized previously~\cite{ambler2017solid}. 
Alternatively, a two-dimensional interface \(h(x,y)\) can first be extracted, after which the fluctuations are averaged along \(y\), equivalently retaining only the \(k_y=0\) modes; this is referred to here as the \emph{2D \(k_y=0\)} approach. 
The distinction between these two procedures is often overlooked: in the former, information along \(y\) is discarded before the interface is identified, whereas in the latter it is retained during interface construction. Supplementary Information S3 presents two-dimensional interfaces obtained using several structural descriptors, including PTM, LOP, the Steinhardt bond-orientational parameter \(q_6\), and the locally averaged Lechner--Dellago parameter \(\bar{q}_6\)~\cite{steinhardt1983bond,lechner2008accurate}. These results illustrate the interfacial fluctuations along the \(y\)-direction and the corresponding 2D \(k_y=0\) analysis procedure.

Finally, for sufficiently thick models, the full two-dimensional fluctuation spectrum was analyzed using nonzero \(k_y\) modes and the interfacial-stiffness tensor in Eq.~\eqref{Eq:stiffness_tensor_fit_equation}; we refer to this as
the \emph{full 2D tensor} approach. The two interface-height fields were Fourier transformed on their native uniform grids without interpolation, and their time-averaged powers were combined before forming the CFM response. All independent modes satisfying \(0.005 < k^2 < 0.03~\mathrm{\AA}^{-2}\) are fitted by unweighted least squares through the origin. One representative of each \(\{\mathbf{k},-\mathbf{k}\}\) pair was retained to avoid double counting.
The fit assumes only the symmetry \(\widetilde{\gamma}_{xy}=\widetilde{\gamma}_{yx}\); no additional crystallographic symmetry constraints or shell averaging were imposed. In particular, \(\widetilde{\gamma}_{xy}\) was fitted freely and was not
constrained to zero.

\subsection*{Interface extraction}
\begin{figure}[t!]
    \centering
    \includegraphics[width=0.75\linewidth]{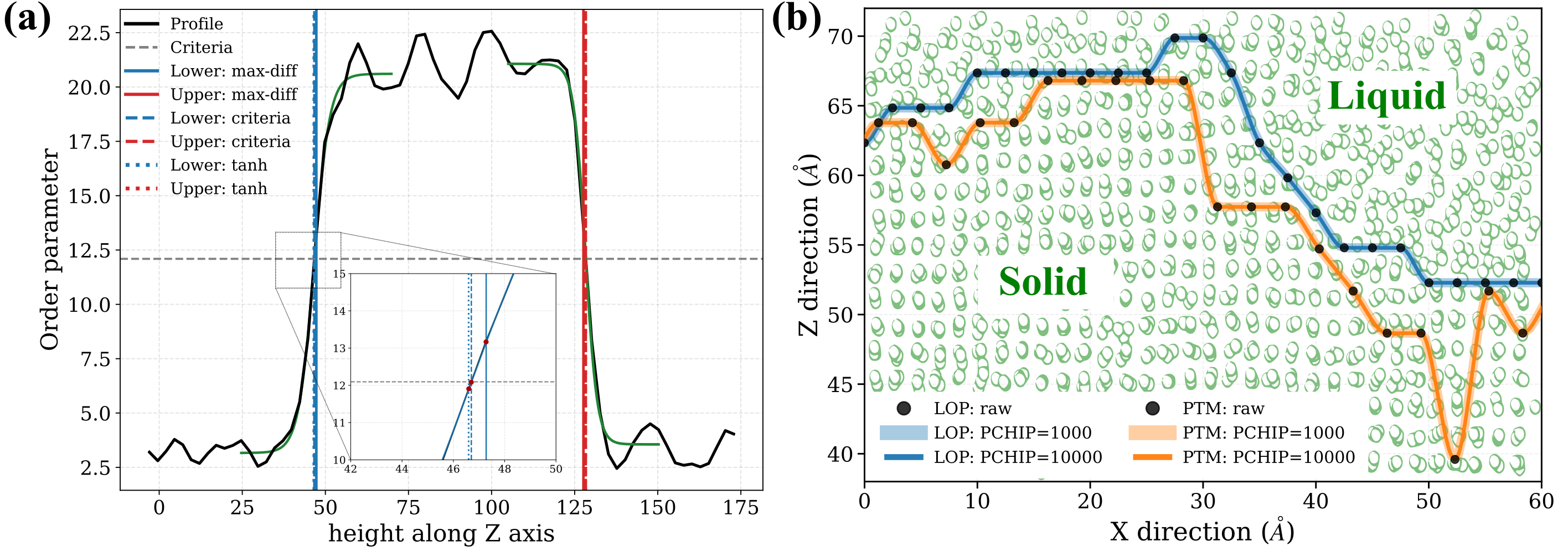}
    \caption{
Effect of interpolation and interface-position criteria on the extracted interface profile.
The figure compares interface positions obtained using different solid--liquid interface identification procedures, including direct grid-based extraction and PCHIP interpolation: representative illustration of (a) interface location determination along the Z axis at a given X at a given time, and (b) resulting reconstructed interface profiles at a given time.
The resulting interface profiles show only minor differences, indicating that the final CFM results are not strongly affected by the interpolation scheme or by small variations in the interface-position criterion.
This test supports the use of a simple and consistent interface extraction procedure in the main analysis.
}
\label{fig:S_interface_extraction}
\end{figure}

Once a coarse-grained scalar field is obtained, the interface position can be extracted as a contour separating solid-like and liquid-like regions. Common approaches include threshold-based isosurface extraction~\cite{hoyt2001method,mishin2014calculation,asadi2015two}, fitting the order-parameter profile to a continuous function such as a hyperbolic tangent profile~\cite{wang2020controlling,haapalehto2022atomistic}, and the maximum-difference method proposed by Brown \textit{et al.}~\cite{brown2019solid}. These methods differ mainly in how the operational interface position is defined: threshold methods require a prescribed isovalue, fitting methods depend on the assumed profile shape and fitting procedure, whereas the maximum-difference method avoids an explicit threshold by locating the position that maximizes the contrast between the two sides of the interface. When the coarse-grained field is sufficiently smooth and the intrinsic interface width is much smaller than the capillary wavelengths retained in the fit, these extraction criteria are expected to yield similar low-\(k\) interface profiles. This is confirmed by our tests, one of which is illustrated in Fig.~\ref{fig:S_interface_extraction}(a), where the three methods give nearly identical interface positions under the present coarse-graining conditions. We therefore use the maximum-difference method in the following analysis because it is straightforward to implement and avoids both a prescribed isovalue and an assumed profile shape. 
Because the influence of the extraction criterion is negligible under the present coarse-graining conditions, it is not treated as a separate uncertainty source in the subsequent analysis.

Piecewise cubic Hermite interpolation (PCHIP) has been used to smooth extracted interface profiles~\cite{brown2019solid}. Its effect is expected to be limited mainly to short-wavelength fluctuations, particularly when the interface has already been densely sampled using kernel smoothing. In our tests, PCHIP shows no influence on the extracted interface profile, as illustrated in Fig.~\ref{fig:S_interface_extraction}(b). Therefore, PCHIP is not used in this work.

\subsection*{Stiffness and anisotropy fitting}
After the SLI stiffness \(\tilde{\gamma}\) has been obtained for each orientation, the anisotropy parameters can be determined by solving the cubic-harmonic stiffness equations~\cite{hoyt2001method,asadi2015two}. For each orientation and fluctuation direction, Eq.~\eqref{Eq:cubic_harmonic_expansion} gives a corresponding expression for the stiffness \(\tilde{\gamma}/\gamma_0\). The measured stiffnesses therefore form a linear system for \(\gamma_0\), \(\varepsilon_1\), and \(\varepsilon_2\).

The sensitivity of this inversion depends on the particular set of orientations included in the fit. Different combinations of stiffness relations lead to linear systems with different condition numbers, and therefore amplify errors in the orientation-resolved stiffnesses to different extents. Consequently, a comparable numerical uncertainty in two stiffness values does not necessarily produce a comparable uncertainty in the fitted anisotropy parameters. This is particularly important for \(\varepsilon_1\) and \(\varepsilon_2\), which are determined largely from differences between stiffnesses of similar magnitude and can therefore be substantially more sensitive than \(\gamma_0\) to small systematic or statistical shifts in individual orientations.

For this reason, uncertainty should be propagated at the anisotropy-fitting level rather than reported only as independent error bars on individual stiffnesses. When redundant orientations are available, they provide an additional consistency check, but only when the corresponding stiffnesses are themselves sufficiently converged. Adding an unconverged orientation can instead increase the sensitivity of the fitted anisotropy parameters.

For sufficiently thick models, the full two-dimensional fluctuation spectrum provides a more constrained determination of the interfacial-stiffness tensor~\cite{du2007properties} by simultaneously incorporating modes with different in-plane wave-vector directions and the symmetry relations of the interface. The resulting stiffness estimates therefore make more complete use of the available fluctuation information than a fit restricted to the \(k_y=0\) modes. In the present work, the anisotropy parameters obtained from the full 2D tensor analysis of the thick models are consequently used as the reference values against which the ribbon-model results are assessed.

\subsection*{Software implementation}\label{sec:package}
To facilitate reproducible implementation of the workflow described here, we developed an open-source Python package for CFM analysis. The package implements the complete post-processing pipeline from atomistic configurations to interfacial stiffnesses and anisotropy parameters, including atom-wise descriptor processing, coarse graining, interface extraction, Fourier analysis, relaxation-time estimation, stiffness fitting, replica-based uncertainty propagation, and diagnostic plotting. It also records the key analysis settings, such as the structural descriptor, coarse-graining parameters, interface-extraction method, fitting \(k\)-window, sampling duration, and replica selection, together with the resulting interfacial properties.

The package is accompanied by example LAMMPS input scripts and analysis notebooks for the pure Al demonstration presented here, providing a practical starting point for applications to other materials systems and interatomic potentials. The source code is available on \href{https://github.com/Kai-Liu-MSE/interface-analyzer}{GitHub}.

\section*{Data availability}\label{sec:data}

The data supporting the findings of this study are available in a Zenodo repository \\ (\href{https://doi.org/10.5281/zenodo.22097543}{https://doi.org/10.5281/zenodo.22097543}). The repository includes representative molecular-dynamics trajectories, simulation inputs and logs, processed interface-position data, and tabulated CFM stiffness and anisotropy results.

\section*{Code availability}\label{sec:code}

The source code, example LAMMPS scripts, and analysis notebooks are available through the interface-analyzer GitHub repository (\href{https://github.com/Kai-Liu-MSE/interface-analyzer}{https://github.com/Kai-Liu-MSE/interface-analyzer}). 
The full MD trajectories generated in this study are large and contain substantial redundancy, and are therefore not included in the public repository. The repository instead contains representative raw trajectory data for testing the CFM analysis workflow, together with the LAMMPS input files, analysis scripts, and configuration files needed to reproduce the simulations and post-processing procedures reported in this work.

\section*{Acknowledgements}\label{sec:ack}

This work was supported by the Spanish Ministry of Science, Innovation, and Universities (MICIU/AEI /10.13039/501100011033) via the KESADIAS project (CNS2024-154608). DES is grateful for support of IMDEA Materials Institute Sabbatical and Mobility Program for External Researchers funded by the State Research Agency of Spain via a Maria de Maeztu Seal of Excellence (CEX2018-000800-M). The computational resources provided by BSC MareNostrum 5 under project ID EHPC-BEN-2025B12-067 are gratefully acknowledged. 

\section*{Author contributions}\label{sec:credit}

{\bf KL}: 
Conceptualization,
Data curation,
Formal analysis,
Investigation,
Methodology,
Software,
Validation,
Visualization,
Writing – original draft.
{\bf DES}: 
Conceptualization,
Formal analysis,
Investigation,
Methodology,
Supervision,
Writing – review and editing.
{\bf DT}: 
Conceptualization,
Formal analysis,
Funding acquisition,
Project administration,
Supervision,
Writing – review and editing.

\section*{Competing interests}

The authors declare no competing interests.

\bibliographystyle{unsrt}
\bibliography{Ref}

\end{document}